\documentclass[acmtog,british,screen=true]{acmart}

\usepackage[T1]{fontenc}
\usepackage[utf8]{inputenc}

\usepackage[british]{babel}

\setcopyright{rightsretained}
\copyrightyear{2024}
\acmYear{2024}
\acmJournal{TOG}
\usepackage{enumitem} % Better control of lists and their spacing

\setlist{nolistsep} % Remove horizontal spacing in lists
\usepackage{caption}
\usepackage{subcaption}
\usepackage{pifont}
\newcommand{\cmark}{\ding{51}}

\usepackage{rotating}

\usepackage[normalem]{ulem} % To allow strikeout text

\newcommand{\rev}[1]{#1} % Hides revisions
\newenvironment{revision}{\par}{\par}

\usepackage{pdfrender}
\newcommand\tablebf[1]{\textpdfrender{TextRenderingMode=FillStroke,LineWidth=0.3}{#1}} % The LineWidth is the boldness

\begin{document}

\title[Evaluating gesture generation in a large-scale open challenge: The GENEA Challenge 2022]{Evaluating gesture generation in a large-scale open challenge:\\The GENEA Challenge 2022}
% \\ Benchmarking the State of the Art


\author{Taras Kucherenko}
\orcid{0000-0001-9838-8848}
\email{tkucherenko@ea.com}
\authornote{Equal contribution and joint first authors.}
\affiliation{
  \institution{SEED -- Electronic Arts (EA)}
  \city{Stockholm}
  \country{Sweden}}

\author{Pieter Wolfert}
\authornotemark[1]
\orcid{0000-0002-7420-7181}
\email{pieter.wolfert@donders.ru.nl}
 \affiliation{%
\institution{%
% Donders Institute for Brain, Cognition \& Behaviour,
Radboud University}
 \city{Nijmegen}
 \country{The Netherlands}}
 \affiliation{%
 \institution{%
% IDLab,
Ghent University – imec}
 \city{Ghent}
 \country{Belgium}
}
 
\author{Youngwoo Yoon}
\authornotemark[1]
\orcid{0000-0003-4286-3421}
\email{youngwoo@etri.re.kr}
\affiliation{%
  \institution{Electronics and Telecommunications Research Institute (ETRI)}
  \city{Daejeon}
  \country{Republic of Korea}
}

\author{Carla Viegas}
\orcid{0000-0003-3385-4101}
\email{cviegas@andrew.cmu.edu}
\affiliation{%
 \institution{Carnegie Mellon University}
 \city{Pittsburgh}
 \country{USA}}
\affiliation{%
 \institution{NOVA University Lisbon}
 \city{Lisbon}
 \country{Portugal}
}
 
\author{Teodor Nikolov}
\orcid{0000-0002-4573-1400}
\email{tnikolov@hotmail.com}
\affiliation{%
 \institution{Ume{\aa} University}
 \city{Ume{\aa}}
 \country{Sweden}
}
\affiliation{%
 \institution{Motorica AB}
 \country{Sweden}
}

\author{Mihail Tsakov}
\orcid{0000-0002-1817-440X}
\email{tsakovm@gmail.com}
\affiliation{%
 \institution{Ume{\aa} University}
 \city{Ume{\aa}}
 \country{Sweden}
}

\author{Gustav Eje Henter}
\orcid{0000-0002-1643-1054}
\email{ghe@kth.se}
\affiliation{%
  \institution{KTH Royal Institute of Technology}
  \city{Stockholm}
  \country{Sweden}
}
\affiliation{%
 \institution{Motorica AB}
 \country{Sweden}
 }

\renewcommand{\shortauthors}{Kucherenko, Wolfert, Yoon et al.}

\begin{abstract}
% IUI abstract draft:
%Co-speech gestures, gestures that accompany speech, play an important role in human communication. Automatic co-speech gesture generation is thus a key enabling technology for embodied conversational agents (ECAs). Research into gesture generation is rapidly gravitating towards data-driven methods. Unfortunately, individual research efforts in the field are difficult to compare: there are no established benchmarks, and each study tends to use its own dataset, motion visualisation, and evaluation methodology. To address this situation, we launched the GENEA Challenges, a series of gesture-generation challenges wherein participating teams built automatic gesture-generation systems on a common dataset, and the resulting systems were evaluated in parallel in a large, crowdsourced user study using the same motion-rendering pipeline. Since differences in evaluation outcomes between systems are solely attributable to differences between the motion-generation methods, this enables benchmarking recent approaches against one another in order to get a better impression of the state of the art in the field. This paper reports on the purpose, design, results, and implications of the second GENEA challenge. 
%Co-speech gesture generation is a key enabling technology for embodied conversational agents.
This paper reports on the second GENEA Challenge to benchmark data-driven automatic co-speech gesture generation. Participating teams used the same speech and motion dataset to build gesture-generation systems. Motion generated by all these systems was rendered to video using a standardised visualisation pipeline and evaluated in several large, crowdsourced user studies. Unlike when comparing different research papers, differences in results are here only due to differences between methods, enabling direct comparison between systems. The dataset was based on 18 hours of full-body motion capture, including fingers, of different persons engaging in a dyadic conversation. Ten teams participated in the challenge across two tiers: full-body and upper-body gesticulation. For each tier, we evaluated both the human-likeness of the gesture motion and its appropriateness for the specific speech signal. Our evaluations decouple human-likeness from gesture appropriateness, which has been a difficult problem in the field.

%The evaluation results are a revolution, and a revelation. 
\rev{The evaluation results show some} synthetic \rev{gesture} conditions \rev{being} rated as significantly more human-like than \rev{3D} human motion capture. To the best of our knowledge, this has not been \rev{demonstrated} before. On the other hand, all synthetic motion is found to be vastly less appropriate for the speech than the original motion-capture recordings. We also find that conventional objective metrics do not correlate well with subjective human-likeness ratings in this large evaluation. The one exception is the Fréchet gesture distance (FGD), which achieves a Kendall's tau rank correlation of around $-0.5$. Based on the challenge results we formulate numerous recommendations for system building and evaluation.
%The main highlight of the results is synthetic motion that is rated as significantly more human-like than the motion from the motion-capture recordings. To the best of our knowledge this has never been shown before on a humanoid mesh avatar. We have also managed to decouple human-likeness from gesture appropriateness in our evaluations, which previously was a major challenge in the field.
%
% Alternative intro and ending for the GENEA Workshop paper abstract:
%Automatic gesture generation is a field of growing interest, and a key technology for enabling embodied conversational agents.
%(...)
%This paper reports on the purpose, design, and of our challenge, with each individual team's entry described in a separate paper also presented at the GENEA Workshop.
%
\end{abstract}

\begin{CCSXML}
<ccs2012>
<concept>
<concept_id>10003120.10003121</concept_id>
<concept_desc>Human-centered computing~Human computer interaction (HCI)</concept_desc>
<concept_significance>500</concept_significance>
</concept>
<concept>
<concept_id>10010147.10010371.10010352</concept_id>
<concept_desc>Computing methodologies~Animation</concept_desc>
<concept_significance>300</concept_significance>
</concept>
</ccs2012>
\end{CCSXML}

\ccsdesc[500]{Human-centered computing~Human computer interaction (HCI)}
\ccsdesc[300]{Computing methodologies~Animation}

%% the work being presented. Separate the keywords with commas.I don't think arXiv offers any way to collect articles into proceedings.
\keywords{animation, gesture generation, embodied conversational agents, evaluation paradigms}

%\settopmatter{printfolios=true} % Turn this on for page numbering on arXiv

\begin{teaserfigure}
  \centering
  \includegraphics[width=\textwidth]{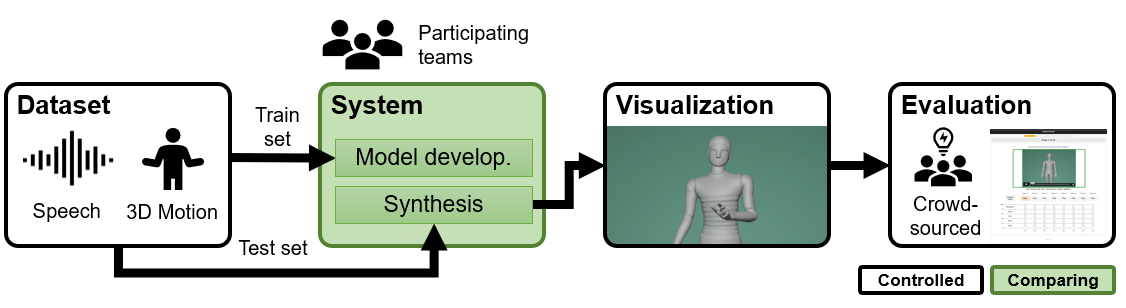}
  \caption{Overview of the GENEA Challenge. We controlled the dataset, visualisation, and evaluation in order to compare different gesture-generation approaches in a fair and systematic way. The dataset includes speech audio, time-aligned speech text transcription, and speaker identity as input modalities and 3D body motion as the output modality. For the synthesised motion from the participating teams, video stimuli were rendered by a shared visualisation pipeline and evaluated jointly in crowdsourced user studies.}
  \label{fig:teaser}
  \Description{The figure shows four boxes with arrows from each one to the next. In order, the boxes are labelled dataset, system, visualization, and evaluation. The dataset box contains speech and 3D motion. The system box involves model development on training data and synthesis using test set inputs, both done by participating teams. This is the only box that differs between teams. The visualization box shows the avatar used and the evaluation is crowdsourced.}
\end{teaserfigure}

\maketitle

\section{Introduction}
% What writing conventions do we use? UK spelling with serial comma? Hyphenation patterns.
This paper is concerned with systems for automatic generation of nonverbal behaviour, and how these can be compared in a fair and systematic way in order to advance the state-of-the-art.
This is of importance as nonverbal behaviour plays a key role in conveying a message in human communication \cite{mcneill1992hand}.
A large part of nonverbal behaviour consists of so called co-speech gestures, spontaneous hand and body gestures that relate closely to the content of the speech \cite{bergmann2011relation}, and that have been shown to improve understanding \cite{holler2018processing}.
Embodied conversational agents (ECAs) benefit from gesticulation, as it improves interaction with social robots \cite{salem2011friendly} and willingness to cooperate with an ECA \cite{salem2013err}. 
Knowledge of how and when to gesture is also needed. This can for example be learnt from interaction data; see, e.g., \citet{jonell2020let}.% and references therein.

Synthetic gestures used to be based on rule-based systems, e.g., \citet{cassell2001beat,salvi2009synface}; see \citet{wagner2014gesture} for a review. These are now being supplanted by data-driven approaches, e.g., \citet{bergmann2009GNetIc,levine2010gesture,chiu2015predicting}, with recent work \cite{yoon2019robots,kucherenko2020gesticulator,alexanderson2020style,yoon2020speech} showing improvements in gesticulation production for ECAs.
For more in-depth reviews of recent data-driven approaches see \citet{liu2021speech,nyatsanga2023comprehensive}.

Unfortunately, results from different gesture-generation studies are typically not directly comparable \cite{wolfert2021review}.
Studies usually rely on different data sources to train their models. 
The visualisations of their generated gestures often have different avatars and production values, which can affect the perception of the gestures.
On top of this, studies make use of a variety of different methodologies to evaluate the gestures. 
All these differences are, however, external to the actual methods that drive the gesture generation. 
%By providing a common dataset for building gesture-generation systems, and common evaluation standards with a common visualisation procedure, one can control for these sources of variation, and enable direct comparison between different methods for co-speech gesture generation.

In this paper, we report on the GENEA\footnote{For ``Generation and Evaluation of Non-verbal Behaviour for Embodied Agents''.} Challenge 2022. %,the second joint gesture-generation challenge.
The aim of the challenge is not to select the best team -- it is not a contest, nor a competition -- but to be able to directly compare different approaches and outcomes.
By providing a common dataset for building gesture-generation systems, along with common evaluation standards and a shared visualisation procedure, we control for all other sources of variation except the system-building itself.
Our setup is unique to the field of gesture generation, making it possible to reliably assess and advance the state of the art, and to measure the gap between it and natural co-speech gestures.
Comparing different methods and their performance also helps identify what matters most in gesture generation, and where the bottlenecks are.
In particular, the results make it abundantly clear that natural-looking data-driven gesture motion is achievable today, but that synthetic gestures are much less appropriate for the accompanying speech than the \rev{natural motion-capture data} is.
The results also show that most objective metrics are not informative about the perceived human-likeness of the generated gestures.

Challenge participants benefit by working on the same problem together with researchers interested in the same topic, strengthening the research community, and get an opportunity to compare their systems to other competitive systems in a large and carefully-executed joint evaluation.
They also work on and contribute towards a standardised evaluation setup which encourages future benchmarking and reproduction of results.
Participants are required to write down their methods, results and experience in a system paper to be presented in conjunction with the challenge itself, giving them a chance to publish their work at ACM ICMI, a leading conference in the field. 
%todo: what did the blizzard challenge achieve? specific info on blizzard challenge is not relevant perse (except the meta stuff)
%
Our concrete contributions are:
\begin{enumerate}
    \item Four large-scale user studies that jointly evaluate a large number of gesture-generation models on a common dataset using a common 3D model and rendering method.
    %\item Demonstrating a new method for subjective assessment of gesture appropriateness for speech, that successfully controls for the human-likeness of the motion.
    \item A subjective evaluation that successfully disentangles motion human-likeness from its appropriateness for the associated speech.
    \item To the best of our knowledge, the first results that identify synthetic gesture motion that surpasses the human-likeness of good \rev{3D} motion-capture data.% on a \rev{humanoid mesh avatar}.
    \item The first clear evidence that synthetic gestures are much less appropriate for the specific speech than natural motion is, even when controlling for the human-likeness of the motion.
    \item A validation study of many objective metrics for predicting motion human-likeness, finding that all metrics except the Fr{\'e}chet gesture distance (FGD) are statistically indistinguishable from chance prediction.
    \item Providing open code and high-quality data in the spirit of open source and reproducible research.
    %to facilitate reproducibility and
    %enabling future research to compare and benchmark against systems from the challenge.
    This includes pre-processed multimodal training, validation, and test datasets; the standardised visualisation; submitted motion and video stimuli; a large number of subjective responses from the studies; and evaluation and analysis code.
    \item Bringing researchers together in order to advance the state of the art in gesture generation, and enabling future research to compare and benchmark against systems and stimuli from the challenge.
\end{enumerate}
%In the long term, we anticipate further benefits in that systems, motion stimuli, and evaluation methods from the challenge can be used as benchmarks in future studies, and in starting to build a database of thoroughly assessed motion stimuli for further research into objective and subjective evaluation.
Materials derived from the challenge are publicly available at \\
\href{https://youngwoo-yoon.github.io/GENEAchallenge2022/}{youngwoo-yoon.github.io/GENEAchallenge2022}.
%, with links to these resources to be added upon paper acceptance.

This paper is an extension of a previously published conference paper on the challenge \cite{yoon2022genea}, adding more comprehensive information and \rev{analyses}, experiments on objective metrics, and a more in-depth discussion of challenge \rev{submissions,} findings, recommendations, and limitations.
The remainder of this paper first (in Sec.\ \ref{sec:background}) briefly discusses current gesture-evaluation practices and how challenges can help overcome their shortcomings.
We then describe the challenge task and dataset in Sec.\ \ref{sec:task_n_data}, followed by a breakdown of the challenge tiers and participating teams in Sec.\ \ref{sec:setup}.
\rev{Sec.\ \ref{sec:systems} then reviews the approaches taken by different challenge entries.}
In Sec.\ \ref{sec:evaluation} we describe the design of the challenge evaluation, with results of the various evaluations recounted in Sec.\ \ref{sec:results} and discussed in Sec.\ \ref{sec:discussion}.
Each of these three sections detail both the core subjective evaluation as well as the objective metrics we computed, in that order.
We round off by discussing challenge limitations (in Sec.\ \ref{sec:limitations}) and summarising its conclusions and implications (in Sec.\ \ref{sec:conclusion}).
%including both objective and subjective evaluation, and the prepared stimuli.
%Sec.\ 6 presents objective and subjective results of the challenge evaluations, including a low-level discussion comparing our multiple conducted user studies.
%Sections 7 and 8 provide a high-level discussion of the challenge, including limitations, implications of our findings, and conclusion.

\section{Related work}
\label{sec:background}
\setcounter{subsection}{1}
\subsubsection{Issues with prior evaluations and evaluation practices}
Most works that propose new gesture-generation methods incorporate an evaluation to support the merits of their method.
Human gesture perception is highly subjective, and there are currently no widely accepted objective measures of gesture perception.
Instead, human assessment via careful user studies is the gold standard in the field.
%Due to the highly subjective aspect of human gestures, many publications have conducted human assessments instead.
However, previous subjective evaluations have several drawbacks, as reviewed in \citet{wolfert2021review}.
Some major issues are the coverage of systems being compared and the scale of the studies.
Like in \citet{sadoughi2019speech,kucherenko2020gesticulator,kucherenko2021moving,alexanderson2020style}, proposed models are at most compared to one or two prior approaches (often a highly similar baseline) or possibly only to ablated versions of the same model.
A large number of studies do not compare their outcomes with other methods at all, let alone other systems trained on the same data.
This creates an insular landscape where particular model families are only applied to particular datasets, and never contrasted against one another.

As for scale, large evaluations are expensive, and studies may not be able to recruit enough participants, thus leaving the differences between many pairs of studied systems unresolved and not statistically significant (cf.\ \citet{yoon2019robots,yoon2020speech}).
%In terms of the study scale, \citet{yoon2019robots} failed to show statistical significance for the majority of pairs of compared systems, due to the low number of evaluation participants.
Questionnaires, which are one popular evaluation methodology (cf.\ \citet{salem2012generation,ishi2018speech,bergmann2010individualized, yoon2019robots, ishii2018generating, shimazu2018generation}) demand a lot of time and cognitive effort from test participants, and may be infeasible to scale up to bigger studies.
In addition, items used in questionnaires differ across studies and the set of questions used is often not standardised.
Evaluations sometimes also fail to anchor system performance against natural (``ground truth'') motion from test data held out from training (e.g., \citet{salem2012generation, ishii2018generating, le2012evaluating}).

% comment : database (e.g., \cite{salem2012generation,ishii2018generating,le2012evaluating})
Studies also differ in the dataset they train on (e.g., \citet{salem2012generation,ishii2018generating,le2012evaluating}) and in how the motion is visualised.
For the latter, some prior work displays motion through stick figures (e.g., \citet{wolfert2019should,kucherenko2019analyzing}), or applies it to a physical agent (e.g., \citet{salem2012generation,ishi2018speech}).
Neither of these may allow the same expressiveness or range of motion as a 3D-rendered \rev{humanoid mesh avatar} as seen in, e.g., \citet{alexanderson2020style,kucherenko2020gesticulator}.

\subsubsection{Benefits of challenges in other fields}
Other fields have done well using challenges to standardise evaluation techniques, establish benchmarks, and track and evolve the state of the art. 
For example, the Blizzard Challenges have since their inception in 2005 (see \citet{black2005blizzard}) helped advance our sister field of text-to-speech (TTS) technology and identified important trends in the specific strengths and weaknesses of different speech-synthesis paradigms \cite{king2014measuring}. 
These challenges are defined by the use of common data and evaluation, and their open participation to both academia and industry.
More specifically, participants are provided a common dataset of speech audio and associated text transcriptions, which they use to build a system that generates synthetic speech audio. 
After the participants submit their systems, the resulting generated speech is subsequently evaluated in a large, joint evaluation, the results of which are provided to the teams.
Submitted entries are identified by anonymised labels in official Blizzard Challenge results, but in practice the vast majority of teams identify which label represents their entry in their paper at the Blizzard Challenge Workshop describing the system that they submitted.
%, yet to the public it is not announced which system is from which team.
Data, evaluation stimuli, and subjective ratings remain available after these challenges, and have been widely used both for benchmarking subsequent TTS systems, e.g., \citet{szekely2012evaluating,charfuelan2013expressive}, and in research on the perception of natural and artificial speech, e.g., \citet{moller2010comparison,yoshimura2016hierarchical,mittag2020deep,saratxaga2016synthetic,govender2019using,huang2022voicemos}.
This has led to the development of new and novel methods, driven by past results, and since participants had access to the same data, significant advances have been made.

% Seems useful for a journal
Challenges are also actively used in the computer-vision community, for instance for benchmarking purposes. Recent NTIRE\ \cite{ntire2020} and CLIC\ \cite{clic2020} challenges, for example, compared systems for image compression and super-resolution respectively, also incorporating subjective human assessments similar to the challenge described in this paper (although NTIRE used a MOS-like setup, which has been found to be less efficient than the side-by-side evaluation methodology we employ \cite{ribeiro2015perceptual}). This addresses the over-reliance on objective metrics in computer-vision evaluation, which, just like in speech quality and gesture generation, do not always align with human perception.
The GENEA Challenge is inspired by these successes of challenges in other fields of study.
%conducted the first challenge in the field of gesture generation.

In 2020 we organised the first gesture-generation challenge, the GENEA Challenge 2020 \cite{kucherenko2021large}.
In addition to being an exercise in benchmarking both new \cite{jinhong_lu_2020_4088376,vladislav_korzun_2020_4088609,thangthai2021speech} and previously-published \cite{alexanderson2020style, kucherenko2019analyzing, yoon2019robots} gesture-generation methods, the results of that challenge have since helped improve gesture-generation benchmarking in other ways as well. Researchers have, for example, used the 2020 visualisation \cite{wang2021integrated,teshima2022deep,zhang2023diffmotion}, and the objective \cite{bhattacharya2021speech2affectivegestures} and subjective \cite{yoon2021sgtoolkit} evaluation methodologies, as a basis for future research. The data has also been used to benchmark subsequent gesture-generation models \cite{ferstl2021expressgesture, yazdian2022gesture2vec}, and even for automatic quality assessment \cite{he2022automatic}.
In this paper, we follow up on the 2020 challenge by reporting on the second gesture-generation challenge, the GENEA Challenge 2022.
This challenge expands the dataset, the range of behaviour considered, and the number of participating teams, and also improves the visualisation and the evaluation practises, in addition to considering objective metrics together with a large subjective evaluation.

% objective metrics
\subsubsection{Objective metrics}
Given that subjective evaluations are labour intensive, time-consuming, and costly, a large number of different objective metrics have been proposed as automated indicators of gesture-generation performance.
Some of these, such as the commonly used average position error (APE) and mean-squared position error (MSE) \cite{wolfert2021review,nyatsanga2023comprehensive}, as well as the ``probability of correct keypoints'' (PCK) and its extension to semantic relevance gesture recall (SRGR) \cite{liu2022beat}, are used to score the similarity of generated poses to a corresponding recording of human motion.
Alternatively, canonical correlation analysis (CCA) can be used to quantify the linear (Pearson) correlations between generated and reference poses \cite{sadoughi2019speech,bozkurt2015affect,jinhong_lu_2020_4088376}.
These methods are likely to struggle with the stochastic, one-to-many nature of human gestures (there is no single ``correct'' way to move), leading to high variance.

To accommodate the stochastic nature of motion, statistics such as the average magnitude of motion acceleration and jerk, and distances between motion speed histograms have been used to quantify how similar generated motion is to the distribution of human motion \cite{kucherenko2021moving}.
%There were a few attempts for evaluating gesture motion objectively.
More recent developments have built on the Fre{\'e}chet inception distance (FID) from image generation \cite{heusel2017gans} to propose new methods for comparing gesture-motion distributions \cite{yoon2020speech, ahuja2020no}.
These methods were later used by, e.g, \citet{ahuja2022low, liu2022learning, ao2022rhythmic, liu2022beat}.
%More recent developments in the same direction, ideas based on the Fréchet distance between distributions of human motion and generated motion have been proposed \cite{yoon2020speech, ahuja2020no} and used in later works \cite{ahuja2022low, liu2022learning, ao2022rhythmic, liu2022beat}.
Beat consistency, which was first proposed for dance motion \cite{li2021ai}, has also been used to assess gesture generation \cite{liu2022learning}.
%was also adopted to evaluate gesture motion . 
However, few of these works experimentally validate their metrics.
In this paper, we use the many conditions and ratings gathered in our user studies to compute and validate five of the above objective metrics for gesture generation.

\section{Task and data}
\label{sec:task_n_data}
The GENEA Challenge 2022 focused on data-driven automatic co-speech gesture generation.
Specifically, given a sequence $\boldsymbol{s}$ of input features that describe human speech -- which could involve any combination of an audio waveform, a time-aligned text transcription, and a speaker ID -- the task is to generate a corresponding sequence $\hat{\boldsymbol{g}}$ of 3D poses describing gesture motion that an agent might perform while uttering this speech (facial expression is not considered).
To enable direct comparison of different data-driven gesture-generation methods, all methods evaluated in the challenge were trained on the same gesture-speech dataset and their motion visualised using the same virtual avatar and rendering pipeline.
This is the same task as in the 2020 challenge, whilst at the same time we changed the dataset (as described below) and refined the evaluation (see Sec.\ \ref{sec:evaluation}).

\subsection{Data}
Compared to 2020, we wanted to expand the dataset to include finger motion, lower-body motion, and material from multiple speakers in dyadic interactions. The latter may provide more natural and interesting gestures than the Trinity Speech-Gesture Dataset \cite{ferstl2018investigating} used in 2020.
We based our challenge on the Talking With Hands 16.2M gesture dataset \cite{lee2019talking}, which comprises 50 hours of audio (captured by close-talking directional microphones) and motion-capture recordings of several pairs of people having a conversation freely on a variety of topics, recorded in distinct takes each about 10 minutes long.
%This is one of the
At the time of the challenge, this was likely the
largest dataset of parallel speech and 3D motion (in joint-angle space) publicly available in the English language.
We removed parts of the dataset (46 out of 116 takes) that lacked audio or had low motion-capture quality, especially for the fingers.
Note that despite the dataset being dyadic by design, the challenge focused on generating one side of the conversation at a time, without awareness of the interaction partner.
The data from the interaction partner in each dyad was typically also included in the challenge material, but as a separate recording without providing links between the two.
This was the case for both the gesture synthesis and for the subsequent evaluation.
%Although recordings from both sides of the dyads was used, it was separated.

\subsubsection{Speech audio and text}
Speech data was shared with participants as WAV audio with no additional processing beyond the anonymisation applied by \citet{lee2019talking}, which replaced many proper nouns with silence. We also provided text transcriptions of the speech, in tab-separated value (TSV) files, and a metadata file with unique anonymous labels for each speaker.
The TSV files were created by first applying \href{https://cloud.google.com/speech-to-text/}{Google Cloud automatic speech recognition} to transcribe the audio recordings, followed by manual review to correct recognition errors and add punctuation for all parts of the dataset (training, validation, and test).

\begin{figure}[!t]
\centering
\includegraphics[width=\columnwidth]{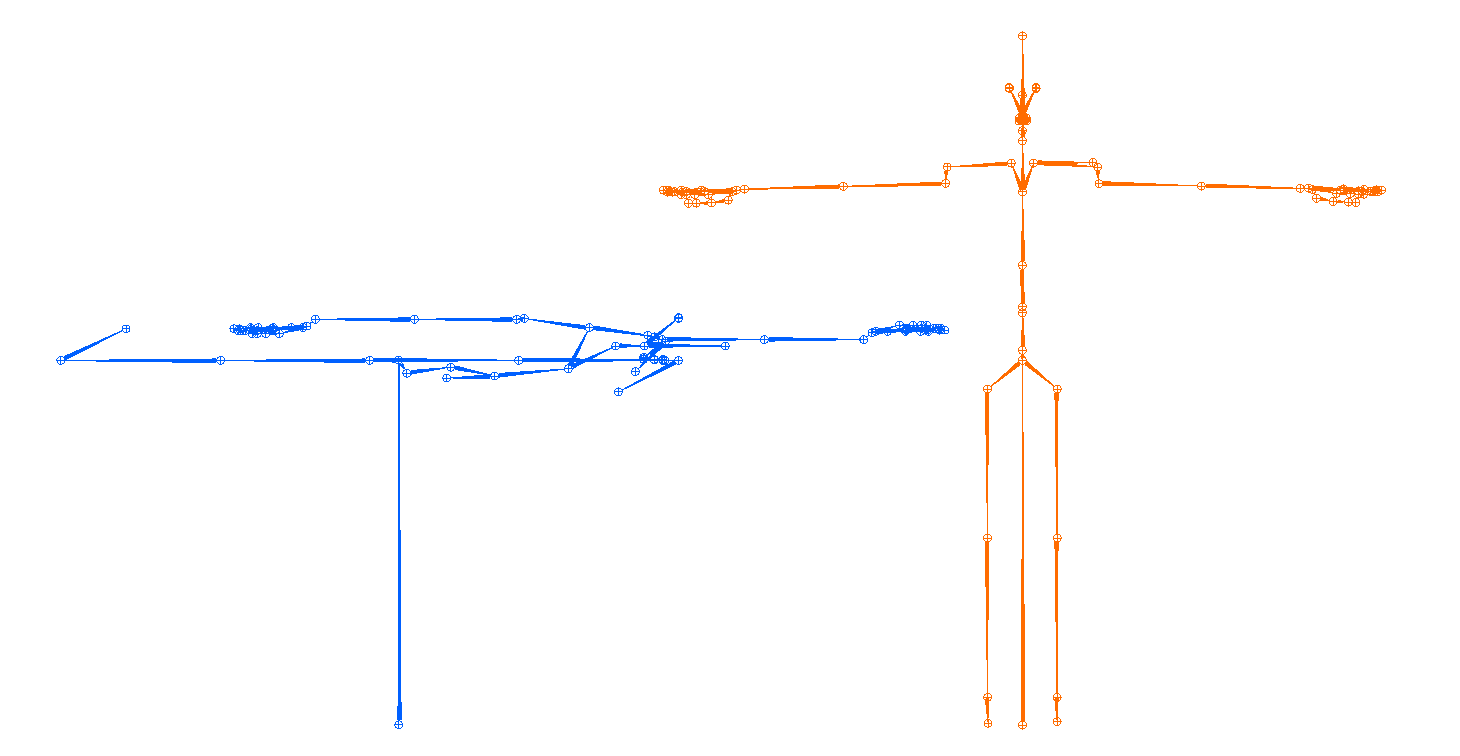}
\caption{Visualisations of the default skeletal pose of the data before and after processing.
On the left (blue) is the original skeletal pose, as found in the Talking With Hands 16.2M dataset shared by \citet{lee2019talking}.
On the right (orange) is the transformed skeletal pose (i.e., T-pose) used for the GENEA Challenge 2022.}
\label{fig:posecomparison}
\vspace{-1\baselineskip}
\Description{The figure shows two human stick figures representing 3D graphics skeletons. The left stick figure is in a very unnatural pose, hovering prone in the air with the joints in an unnatural tangle. No human body will be able to assume that pose without severe injury. The right stick figure shows the same skeleton, but in a standard T-pose, with the arms straight out and legs straight down and together.}
\end{figure}

\subsubsection{Motion data}
Motion data was downsampled from 120 to 30 frames per second and further transformed in two ways:
%\begin{enumerate}
    %\item
    
    \rev{First,} we updated the default skeleton relative to which all motion data is defined, away from what appeared to be a contorted and arbitrary definition to a standard ``T-pose'', as shown in Fig.\ \ref{fig:posecomparison}.
    The T-pose is an animation-industry standard wherein all joint rotation values are described in relation to a T-shaped skeleton.
    This standard is
    %recognised across the animation industry and is
    widely adopted by existing 3D digital content-creation software like Blender and Maya.
    In fact, it is often the required 3D-skeleton pose when transferring the motion of one character onto another during animation re-targeting.
    Furthermore, the T-pose is expected to have better mathematical properties due to its symmetry and shape.
    In particular, the pose more closely resembles the poses found in the motion-capture data.
    As a result of this, most of the joint rotation values are expected to be closely distributed around zero.
    Consequently, this would reduce the risk of phase wrapping and gimbal locking across the skeleton, lending itself to smoother behaviour and interpolation in the Euler-angle space.
    This in turn leads to data that is more numerically stable, making it more practical for training machine-learning models.
    The data was recomputed to match a T-pose using motion re-targeting inside MotionBuilder
    (code and documentation are available\footnote{\href{https://github.com/TeoNikolov/genea_visualizer/tree/master/scripts}{github.com/TeoNikolov/genea\_visualizer/tree/master/scripts}})%(documentation can be found at \href{https://github.com/TeoNikolov/genea_visualizer/tree/master/scripts}{https://github.com/TeoNikolov/genea\_visualizer/tree/master/scripts})
    , retaining as much of the original visual quality as possible, whilst ensuring that the data had no discontinuities (e.g., at rotations near $180^{\circ}$).
    We found that this transformation substantially improved the output of the baseline system UBA in Sec.\ \ref{ssec:teams}.
    %\item
    
    \rev{Second,} we standardised the position and orientation of speakers in all takes.
    Originally, each take would have the two speakers occupy two locations and face each other.
    We standardised this on a per-take basis such that all speakers, on average, face the same direction, and occupy the same location.
    More technically, in a right-hand $xyz$ Cartesian coordinate system ($y$-up, $z$-forward), each speaker is on average positioned at world origin ($[{x=0}$, ${y=0}$, ${z=0}]$), and on average facing the positive $z$-axis (a directional vector $[{x=0}$, ${y=0}$, ${z=1}]$).
    Averaging was done for each take separately after taking 250 equidistant samples of the hips position and orientation, and then using linear-algebra operations to correct for the difference between the original and the standardised values.
    This change was made to streamline data visualisation and to remove potential confusion due to different absolute positions and orientations across different takes.
%\end{enumerate}
The transformed motion data was shared with participants in the Biovision Hierarchy (BVH) format.

\subsubsection{Data splits}
The challenge data was split into a training set (18 h), a validation set (40 min), and a test set (40 min), with only the training and validation sets initially shared with participating teams.
%All these data subsets have been made publicly available at \href{https://zenodo.org/doi/10.5281/zenodo.6998230}{https://zenodo.org/doi/10.5281/zenodo.6998230} after the challenge, and have been downloaded over 500 times when this article went to press.
The validation and test data each comprised 40 \emph{chunks} (contiguous excerpts approximately one minute long), to promote generation methods that are stable over long segments of speech, and was restricted to takes with finger-motion tracking for the chosen speaker.
Some recordings with finger-capture data were excluded from consideration due to poor motion-capture quality, based on visual inspection of a short sample from each recording.
The validation data was intended for internal benchmarking during development, so participants were allowed to train their final submitted models on both training and validation data if they wished.
After the challenge, all the data subsets were made publicly available at \href{https://zenodo.org/doi/10.5281/zenodo.6998230}{zenodo.org/doi/10.5281/zenodo.6998230}, and had been downloaded over 500 times when this article went to press.

\subsubsection{Usage rules}
Teams were allowed to only train on a subset of the data and were allowed to enhance the data they trained on however they liked, for instance by manual annotation or by post-processing the speech and the motion.
%and/or by only training on a subset of the data.
They were also allowed to make use of additional speech data (audio and text) from other sources, and models derived from such data, e.g., BERT \citep{devlin2018bert} and wav2vec \citep{wav2vec2020}.
However, it was not permitted to use any other motion data, nor any pre-trained motion models, other than what the organisers provided for the challenge.
Otherwise, the result would be likely to strongly depend on how much data each team can get access to (as has been the case in many Blizzard Challenges in speech synthesis), which is not an interesting scientific conclusion to replicate.

\section{Setup and participation}
\label{sec:setup}
The challenge began on May 16, 2022, when speech-motion training data was released to participating teams.
Test inputs (WAV, TSV, and speaker ID, but no motion output) were released to the teams on June 20, with teams required to submit BVH files with their generated gesture motion for these inputs by June 27,
%The BVH format is the same
in the same format as that used by the challenge training data.
Manual tweaking of test inputs or the output motion was not allowed, since the idea was to evaluate synthesis performance in an unattended setting.
As a precondition for participating in the evaluation, teams agreed to submit a companion paper describing their system for review and possible publication at the conference where the challenge took place.
Evaluations took place only after the generated motion was submitted by all teams.

\subsection{Tiers}
The challenge evaluation was divided into two tiers, one for full-body motion and one for upper-body motion only.
Each tier had its own reasons for being included.
On the one hand, the data comprises recorded full-body motion from conversational interactions.
It can furthermore be argued that human embodied conversation uses the full body.
Also, generating full-body behaviour seems like a harder problem, since it represents a higher-dimensional probability distribution which is more difficult to learn from a statistical perspective.
Therefore, if full-body generation is solved, restricted versions of the problem can be expected to be solved as well.
On the other hand, it is debatable to what extent the motion of the lower body whilst speaking constitutes co-speech gestures that depend on the speech, over other aspects such as proxemics and stance in response to the other parties in a conversation (which is data that was not provided to challenge participants this time).
As a result, including lower-body motion may add unnecessary complexity to the gesture generation problem, and act as a distraction when evaluating the quality of generated gestures.
%In addition, f
Focusing on the upper body also is more consistent with earlier evaluations of co-speech gesture generation, such as the GENEA Challenge 2020 \cite{kucherenko2021large}.
Because it is not clear which perspective to apply, our evaluation included one tier each for full-body and upper-body motion.
Teams could enter motion into either tier, or into both, but could only make one submission per tier.
Teams that entered into both tiers were allowed to submit different motion (BVH files) to each tier, if they wished.
Both tiers used the same training data but differed in which parts of the avatar that were allowed to move, and in the camera angle used for the video stimuli in the evaluation, as follows:
\begin{description}
\item[Full-body tier] In this tier, the entire virtual character was free to move, including moving around in space relative to the fixed camera.
Motion was visualised from an angle facing the character that showed most of the legs, but not where the feet touched the ground.
This perspective was chosen to show as much as possible of the character, whilst obscuring foot penetration or foot sliding artefacts from view, since these artefacts arguably do not relate to co-speech gestures, yet they may influence ratings if visible.
An example of this camera perspective can be seen in Fig.\ \ref{fig:hemvipgui}.
\item[Upper-body tier] In this tier, the virtual character used a fixed position and a fixed pose from the hips down, with only the upper body free to move.
Motion was visualised from a camera angle facing the character, cropped slightly below the hips, such that the hands always should remain in view.
Any motion of the lower-body joints in submitted BVH files was ignored by the visualisation.
This camera perspective is shown in Fig.\ \ref{fig:evaluation_interface_pairwise}.
\end{description}%
\begin{table*}
%\small
\centering
\caption{Conditions participating in the evaluation. Conditions are ordered based on their median human-likeness scores from higher to lower (see Table \ref{tab:stats})\rev{; this is why the previously published baselines appear near the bottom of the list, but not at the very end}.
The following abbreviations were used:
\textit{AR} for ``Auto-regression'',
\textit{CNN} for ``Convolutional Neural Network'',
\textit{RNN} for ``Recurrent Neural Network'',
\textit{SA} for ``Neural self-attention'' (e.g., Transformers), \textit{GANs} for ``Generative adversarial networks or adversarial loss terms'',
\textit{VAEs} for ``Variational auto-encoders'',
\textit{MGs} for ``Motion graphs'',
\textit{Frame-wise} for ``Generating output frame-by-frame'', \textit{Stoch.\ output} for ``Stochastic output'' (different output possible even if the inputs are the same),
and \textit{Smoothed} for ``Smoothing was applied''.}
\small%
\begin{tabular}{@{}lll|ccc|c|ccccl|ccc@{}}
\toprule 
 & \multicolumn{2}{c|}{Per-tier} & \multicolumn{3}{c|}{Inputs used} & Hands & \multicolumn{5}{c|}{Techniques used} & Frame- & Stoch. & Smoo- \tabularnewline
Baseline or team name & \multicolumn{2}{c|}{label} & Aud. & Text & Sp. ID & fixed & AR & RNNs & SA & VAEs & Other & wise & output & thed \tabularnewline
%\midrule
%Natural motion & \cmark & \cmark & \cmark & \multicolumn{7}{c@{}}{Not relevant} \tabularnewline
\midrule 
GestureMaster \rev{[\citeauthor{zhou2022gesturemaster}]} & FSA & USQ & \cmark & \cmark & \cmark &  & &  &  &  & Hand-crafted rules, MGs &  &  & \cmark\tabularnewline
Forgerons \rev{[\citeauthor{ghorbani2022exemplar}]} & FSC & USO & \cmark &  &  & \cmark & \cmark & \cmark &  & \cmark &  & \cmark & \cmark &   \tabularnewline
DeepMotion \rev{[\citeauthor{lu2022deepmotion}]} & FSI & USJ & \cmark & \cmark &  & \cmark & \cmark &  & \cmark & \cmark & CNNs & \cmark & \cmark & \tabularnewline
DSI \rev{[\citeauthor{saleh2022hybrid}]} & FSF &  & \cmark &  &  &  & \cmark & \cmark & \cmark &  &  &  &  & \tabularnewline
UEA Digital Humans \rev{[\citeauthor{windle2022uea}]} & FSG & USM & \cmark & \cmark & \cmark &  & & \cmark &  &  &  & \cmark &  & \tabularnewline
ReprGesture \rev{[\citeauthor{yang2022reprgesture}]} & & USN & \cmark & \cmark & \cmark & \cmark & \cmark & \cmark & \cmark & \cmark & CNNs, GANs &  &  & \cmark\tabularnewline
IVI Lab \rev{[\citeauthor{chang2022ivi}]} & FSH & USK & \cmark & \cmark & \cmark & \cmark & \cmark & \cmark &  &  &  & \cmark & \cmark & \cmark\tabularnewline
FineMotion \rev{[\citeauthor{korzun2022recell}]} & FSD & & \cmark & \cmark &  &  & \cmark & \cmark &  &  &  & \cmark &  & \cmark\tabularnewline
Murple AI lab \rev{(no paper submitted)} & \multicolumn{2}{c|}{Not revealed} & \cmark &  &  &  & \cmark & \cmark &  &  & Normalising flows & \cmark & \cmark & 
\tabularnewline
Text-only baseline
%[\citeauthor{yoon2019robots}]
& FBT & UBT & & \cmark & & \cmark & \cmark & \cmark  &  &  &  & \cmark &  & \cmark \tabularnewline
Audio-only baseline
%[\citeauthor{kucherenko2019analyzing}]
& & UBA & \cmark &  &  & &  &  \cmark &  &  &  & \cmark &  & \cmark \tabularnewline
%\midrule 
TransGesture \rev{[\citeauthor{kaneko2022transgesture}]} & & USL & \cmark &  & & \cmark  & \cmark & \cmark &  &  &  & \cmark &  & \cmark\tabularnewline
\bottomrule
\end{tabular}
\label{tab:conditions}
\end{table*}

\subsection{Baselines and participating teams}
\label{ssec:teams}
The challenge evaluation featured three types of motion sources: natural motion capture from the speakers in the database, baseline systems based on open code, and submissions by teams participating in the challenge.
We call each source of motion in a tier a \emph{condition} (not a ``system'', since not all conditions represent motion synthesised by an artificial system).
Each condition was assigned a unique three-letter \emph{label} or \emph{condition ID}, where the first character signifies the tier, with F for the full-body tier and U for the upper-body tier.

Natural motion was labelled \textbf{FNA} in the full-body tier and \textbf{UNA} in the upper-body tier (NA for ``natural'').
These conditions can be seen as a top line, and surpassing their performance essentially means outperforming the dataset itself, subject to limitations due to the motion capture and visualisation (discussed in Secs.\ \ref{ssec:humlikecomments} and \ref{sec:limitations}).

The natural top line can be contrasted against the two baseline systems included in the challenge, which represent previously published gesture-generation methods \rev{that have been adapted to run on the 2022 challenge training data.
The systems were selected with the requirements to be (1) open-sourced and well documented, and that (2) their authors were available to adapt the methods to the new data and also help challenge organisers and participants, should any issues arise.
Unfortunately, none of the top-performing models from the previous challenge satisfied both conditions, whereas the two 2020 baselines did.
Adapting these baselines to the present challenge provides continuity with the previous iteration of the challenge and helps track the progress of the field.}
\rev{The} two baselines \rev{were thus}:
\begin{description}
\item [Text-based baseline (FBT/UBT)]
This motion was generated by the gesture-synthesis approach from \citet{yoon2019robots} (which takes text transcriptions with word-level timestamps as the input) but adapted to joint rotations. % as described in \cite{kucherenko2021large}.
A neural sequence-to-sequence architecture is used, where an encoder processes a sequence of speech words and a decoder outputs a sequence of human poses.
Motion from this baseline used a fixed lower body but was included in both tiers, as conditions \textbf{FBT} and \textbf{UBT}
(B for ``baseline'' and T for ``text'').
The code is publicly available online at \href{https://github.com/youngwoo-yoon/Co-Speech_Gesture_Generation}{github.com/youngwoo-yoon/Co-Speech\_Gesture\_Generation}.
\item [Audio-based baseline (UBA)]
This motion was generated by the Audio2Repr2Pose motion-synthesis approach of \citet{kucherenko2019analyzing}, which only takes speech audio into account when generating output, adapted to joint rotations. % as described in \cite{kucherenko2021large}.
This model uses a chain of two neural networks: one maps from speech to pose representation and another decodes the representation to a pose, generating motion frame-by-frame by sliding a window over the speech input.
Motion from this baseline was only included in the upper-body tier, as condition \textbf{UBA} (A for ``audio'').
The code is publicly available online at \href{https://github.com/genea-workshop/Speech_driven_gesture_generation_with_autoencoder/tree/GENEA_2022/}{github.com/genea-workshop/Speech\_driven\_gesture\_
generation\_with\_autoencoder}.
\end{description}
\rev{Source code for replicating the two baselines was available to participating teams during the challenge.}

Separate from top lines and baselines, a total of 10 teams participated in the GENEA evaluation, with 8 \emph{entries} (a.k.a.\ \emph{submissions}) to the full-body tier and 8 entries to the upper-body tier.
\rev{Together with test-set mocap and the baselines, this makes a total of 10 conditions in the full-body tier and 11 in the upper-body tier.}
Submissions were labelled with the prefix FS and US (S for ``submission'') depending on the tier, followed by a single character to distinguish between different submissions in the same tier.
%\rev{Note that the number of labels include the baseline systems, but exclude labels for Murple AI lab.}
In particular, challenge entries to the full-body tier were labelled \textbf{FSA}--\textbf{FSI}, and entries to the upper-body tier were labelled \textbf{USJ}--\textbf{USQ}.
Condition FSE was withdrawn before the evaluation.
These labels are anonymous and have no relationship to team identities, but teams were free to reveal their label(s) in papers describing their systems, if they wished.

Table \ref{tab:conditions} lists the baselines and participating teams, with basic information about their respective approaches \rev{and references to their system-description papers}.
%\rev{Paper meta-information has temporarily been omitted to preserve anonymity during review.}
All teams but one published a paper about their system, and all of the published papers chose to reveal the label(s) of their submitted systems.
We have therefore included that label information in Table \ref{tab:conditions}.
%One team lacks information, since they did not submit a paper for review.

\begin{revision}
%\section{Participating systems}
\section{Methods used by challenge entries}
\label{sec:systems}
Based on the publications referenced in Table \ref{tab:conditions}, we now review the technical approaches taken by the different teams this challenge, and (in Sec.\ \ref{ssec:2020challenge}) contrast them against the five 2020 challenge entries.
Note that we do not discuss the Murple AI Lab submission, since that team did not submit a system-description paper.

\subsection{Motion-generation approaches in 2022}
%We review the participating systems in terms of their technical approaches. 
Most of the teams proposed neural network-based methods for generating pose sequences.
The RNN and Transformer architectures were the most common choices, which effectively led to smaller architectural differences between the systems.
The GestureMaster team was unique among the teams in utilising motion matching \citep{buttner2015motion}, which involves extracting and combining snippets of motion based on the training dataset, for their approach.
Using ChoreoMaster \citep{chen2021choreomaster} for dance as a starting point, they present a motion graph-based matching method for optimally selecting and combining gesture motion clips into a sequence, based on three criteria: rhythm, style, and the transition between consecutive clips \cite{zhou2022gesturemaster}.
To generate the target style embeddings, they fed speech audio into a trained StyleGestures \citep{alexanderson2020style} system.

The other approaches that employed neural networks presented various variations on input-feature context encoding and output-feature decoding for gesture generation.
%Note that we do not discuss the Murple AI Lab submission, since they did not submit a system-description paper.

%First, there are two systems that made changes to the attention mechanism that is part of their Sequence-to-Sequence approach. For example, the TransGesture team \citep{kaneko2022transgesture} employed RNN-Transducers to concentrate solely on past information during the generation phase, enabling the utilisation of an ongoing audio stream for gesture generation. Additionally, the IVI team \citep{chang2022ivi} integrated the Tacotron2 concept (\citep{shen2018natural}) and implemented locally constrained attention to synchronise motion with input speech temporally.

%Next, two systems introduced changes to the decoding structure.
%Two teams focussed on modifying the decoder.
%The FineMotion team \citep{korzun2022recell} proposed a linear layer-based decoder utilising the previous frame and speech context as input, instead of an RNN-based decoder, in order to improve motion stability between frames.
%The UEA Digital Humans submission \citep{windle2022uea} made use of a combination of decoders, where each individual decoder would generate parts of the resulting pose (face, upper-body, and hands). 

To begin with, several approaches made use of custom representation learning for the different modalities.
In particular, both the DeepMotion \citep{lu2022deepmotion} and ReprGesture teams \citep{yang2022reprgesture} created pre-trained modality representations for their submissions.
DeepMotion used a VQ-VAE to map motion data into a discrete space, whilst the ReprGesture team performed both modality-invariant and modality-specific representation learning, combining the two types of features for gesture generation.
In a related move, the Forgerons team \citep{ghorbani2022exemplar} introduced a style-encoding component to learn to encode the style of an input motion.
%that utilises motion samples as input

Two submissions used sequence-to-sequence models with variants of neural attention mechanisms.
The TransGesture submission \citep{kaneko2022transgesture} employed RNN-Transducers that only make use of past information during synthesis, meaning that the approach can be applied to streaming audio with no algorithmic latency.
The IVI Lab submission \citep{chang2022ivi} was based on the Tacotron 2 text-to-speech architecture \citep{shen2018natural}, but modified to use locally constrained attention when synchronising the motion with the input speech audio.

The DSI submission \citep{saleh2022hybrid}, similar to earlier work in autoregressive gesture-generation \citep{kucherenko2020gesticulator}, employed curriculum learning to reduce the error accumulation inherent in autoregressive generation. 

Finally, two teams focussed on modifying the decoder.
The FineMotion team \citep{korzun2022recell} proposed a linear layer-based decoder utilising the previous frame and speech context as input, instead of an RNN-based decoder, to improve motion stability between frames.
The UEA Digital Humans submission \citep{windle2022uea} made use of a combination of decoders, where each individual decoder would generate parts of the resulting pose (face, upper-body, and hands). 

In the next few subsections, we delve deeper into the representation and processing of the input and output data performed by various teams, seeing that these aspects can have a major impact on the results produced by data-driven synthesis methods.
%We delve deeper into the analysis of the systems, specifically examining the aspects outlined below.
In Sec.\ \ref{subsec:system_analysis} -- after reporting on the challenge results -- we draw lessons from the system performance in relation to these aspects.

\subsubsection{Input modalities and their representation}
As shown in Table~\ref{tab:conditions}, all submissions from participating teams used speech audio input, with some additionally employing speech text. 
Strictly speaking, speech text is not independent from speech audio, since speech audio carries all the information provided by the speech text.
(Indeed, the text was derived by transcribing the audio recordings.)
However, speech audio exposes rhythmic and paralinguistic information, whereas speech text offers a more direct representation of speech lexical content.
This makes these different representations suitable for different tasks relevant to gesture generation \citep{kucherenko2022multimodal}.
Hence, it is reasonable for a submission to utilise both audio and text.
Additionally, to enable systems to capture individual differences in gesturing behaviour, a few teams made use of the provided speaker ID information as input.

For speech audio, most teams (4 out of 9) relied on mel-frequency cepstrum coefficient (MFCC) features \citep{davis1980comparison}.
Some used pre-trained off-the-shelf foundation models to represent the input modality.
In particular, the ReprGesture submission used WavLM \citep{chen2022wavlm} and the UEA submission used PASE+ \citep{ravanelli2020multi} for encoding the audio input.
Among the submissions that used speech text as an input, most employed FastText to provide word embeddings \citep{bojanowski2017enriching}.
Four teams used speaker ID as input. 
Some used one-hot vectors of speaker ID as input features \cite{chang2022ivi, windle2022uea}, whilst the Forgerons submission \cite{ghorbani2022exemplar} implemented methods for style control based on a given motion exemplar.

\subsubsection{Output motion representation} 
For the challenge evaluation, teams had to generate BVH files, the same format as used to distribute the dataset.
Poses in these BVH files are represented by root-node positions and Euler angles for joint rotations. 
Due to discontinuities in Euler angles representations \citep{zhou2019continuity}, no team trained their neural networks to output Euler angles directly.
Instead, the IVI Lab \citep{chang2022ivi} and the TransGesture \citep{kaneko2022transgesture} submissions used exponential map representations for the output motion \citep{grassia1998practical}, whereas the DeepMotion \citep{lu2022deepmotion} and the UEA \citep{windle2022uea} submissions used a 6-dimensional representation \citep{zhou2019continuity}.
The DSI \citep{saleh2022hybrid} and the ReprGesture \citep{yang2022reprgesture} teams utilised rotation matrices, whilst the FineMotion \citep{korzun2022recell} team used an axis-angle representation.
The Forgerons team \citep{ghorbani2022exemplar} employed a 2-axis rotation matrix \citep{zhang2018mode} to represent joint rotations and used a mixture of joint position, rotation, positional velocity, and rotational velocity as the output data of the gesture synthesis model. 

\subsubsection{Pre- and post-processing.} 
One popular strategy for training data pre-processing was to exclude segments where the character was not speaking,
%non-speaking segments
which in theory would make some methods produce better co-speech gesture models, as argued by the DeepMotion \citep{lu2022deepmotion} and Forgerons \citep{ghorbani2022exemplar} submissions. 
The ReprGesture team \citep{yang2022reprgesture} decided to use data from only one speaker in the training set due to the potential interference caused by style diversity among speakers.

The most common post-processing technique among the submissions was to apply smoothing to the raw output motion.
For systems that generate poses frame by frame, applying a smoothing filter often helps reduce visual artefacts in case of discontinuous, jerky, or jittery motion in the original model output.

% \vspace{0.5em}
% \noindent \textbf{Something else?} ...

%\subsection{Comparison to the first challenge}
\subsection{Motion-generation approaches in 2020}
\label{ssec:2020challenge}
For the first GENEA Challenge \citep{kucherenko2021large}, in 2020, five teams (\citet{jinhong_lu_2020_4088376,simon_alexanderson_2020_4088600,vladislav_korzun_2020_4088609,kunkun_pang_2020_4090879,thangthai2021speech}) submitted entries to the crowd-sourced evaluation.
There are some key differences between the systems that were entered into the first challenge and the more recent challenge.
Some of these differences were due to the challenges setup.
All submissions the first challenge only supported gesture generation for one individual, since the dataset of that challenge only was sourced from one single person.
In addition, the first challenge only considered upper-body motion.

All submissions to the first challenge used speech audio as input, represented using MFCCs, with some teams also making use of the provided text transcriptions.
%For text-dericed features, three of the five teams relied on learnt text embeddings
The three submissions that made use of text transcripts used learnt embeddings
%, chiefly fastText \citep{bojanowski2017enriching}.
like BERT \citep{devlin2018bert} and GloVe \citep{pennington2014glove} to extract and represent information from the text.
As for the output poses, three teams used 3D joint rotations as the output features for the model to learn, with two teams instead relying on an exponential map representation.
Two approaches relied on autoregressive architectures \cite{simon_alexanderson_2020_4088600, kunkun_pang_2020_4090879}, whilst the other either relied on RNNs \cite{vladislav_korzun_2020_4088609} or an autoencoder \cite{thangthai2021speech, jinhong_lu_2020_4088376}.
Importantly, unlike 2022, none of the 2020 entries used motion graphs for output generation.

We will return to the different approaches taken by participating teams in 2020 and 2022 in Sec.\ \ref{ssec:whatlearnt}, and relate the approaches to the outcomes of the respective challenges, after having reported on the 2022 challenge evaluation and its results.

\end{revision}

\section{Evaluation}
\label{sec:evaluation}
We conducted a large-scale, crowdsourced, joint evaluation of gesture motion from the 10 full-body conditions and 11 upper-body conditions (listed in Table \ref{tab:conditions}) in parallel using a within-subject design (i.e., every rater was exposed to and evaluated all conditions in each tier).
%The evaluation focused on gesture quality of the various submitted systems.
The systems were evaluated in terms of the human-likeness of the gesture motion itself, as well as the appropriateness (a.k.a.\ specificity) of the gestures for the given input speech.
The central difference from other gesture-generation evaluations is that all systems in our evaluation used the same motion data, the same visualisation/embodiment, and were rated together using the same evaluation methodology; only the motion-generation systems differed between the different entries that were compared. This allows the performance of systems to be compared directly, and the design aspects that influence performance can be traced more efficiently than in most previous publications.
%\citet{jonell2020iva_crowd} recently found that the results from crowdsourcing evaluations were not significantly different from in-lab evaluations in terms of results and consistency.
%We therefore adopted an entirely crowdsourced approach, as opposed to, for example, the Blizzard Challenge, which has used a mixed approach.
The subjective evaluation used an entirely crowdsourced approach, with attention checks used to exclude participants that were not paying attention, as detailed in Sec.\ \ref{subs:att_checks_hl}.
The remainder of this section describes the experiments we performed.
Results of the subjective evaluation are subsequently presented in Sec.\ \ref{sec:results} and discussed in Sec.\ \ref{sec:discussion}.

Although the aim of the challenge is to quantify how natural and appropriate motion appears to human observers, we have also seized the opportunity to compute a number of objective metrics of motion quality on the motion materials in the evaluation.
The design of that experiment is described in Sec.\ \ref{ssec:obj_metrics}, with results reported in Sec.\ \ref{ssec:objectiveresults} and discussed in Sec.\ \ref{ssec:obj_discussion}.
We see this primarily as an evaluation of the metrics themselves, and not as an evaluation of the different conditions in the challenge.
%challenge computes and provides some objective metrics of motion quality (see Sec.\ \ref{ssec:obj_metrics}), the focus of the GENEA Challenge is on a large-scale, crowdsourced subjective evaluation of the generated gestures.

\subsection{Subjective evaluation design philosophy}
For each tier, two \rev{different} aspects of the generated gestures were evaluated (with one study per aspect and tier):
\begin{description}
\item[Human-likeness] Whether the motion of the virtual character \rev{visually} looks like the motion of a real human, controlling for the effect of the speech. We sometimes use ``motion quality'' as a synonym for this.
\item[Appropriateness] (a.k.a.\ ``specificity'') Whether the motion of the virtual character is appropriate for the given speech, controlling for the human-likeness of the motion.
\end{description}
\rev{Human-likeness is thus a unimodal and unconditional quality measure (it only depends on the output motion), whereas speech appropriateness is multimodal and conditional on the speech.
The former assesses system output quality whilst the latter assesses how well the output of the system relates to its input, disregarding the intrinsic quality of the output as much as possible.}
\rev{A deeper motivation for separating conditional and unconditional evaluation follows, with} more details about the \rev{two different} evaluations provided in Sections \ref{sec:humlike} and \ref{sec:approp} \rev{further below}.

\subsubsection{Why separate conditional and unconditional performance measures?}
%\rev{Human-likeness is thus an unconditional quality measure (it only depends the output motion), whereas appropriateness is conditional on the speech, and assesses how well the output of the system relates to its input.
%This separation of conditional and unconditional evaluation measures recurs . It is possible to do better on one and.}
\begin{revision}
The complementarity of conditional and unconditional performance measures has long been recognised in other fields, and our decision to perform both unconditional (unimodal) and conditional (multimodal) subjective evaluations reflects a widespread distinction seen in both objective as well as subjective evaluation of synthesis methods in general.
%The objective evaluation metrics we consider in Sec.\ \ref{ssec:obj_metrics} exhibit a similar dichotomy, where most (like our human-likeness evaluation) are insensitive to how the generated output motion sequences are paired with speech, whereas one of them (CCA) involves an element of appropriateness, since it .
In image generation from text prompts, the Fr\'{e}chet inception distance (FID) \citep{heusel2017gans} is a widely used unconditional metric of synthesis quality.
It does not take the input text into account at all, and would not notice disconnects between the input and output modalities, such as if the prompt ``elephant'' were to generate an image of an ant and vice versa.
In contrast, multimodal CLIP embeddings \citep{radford2021learning} can assess the extent to which a synthetic image matches its corresponding text prompt, regardless of visual quality.

Similarly, our sister field of text-to-speech distinguishes between the concepts of quality (or naturalness) and intelligibility, which are closely related to our respective constructs of human-likeness and appropriateness.
Speech quality is usually assessed in a unimodal fashion (only involving audio), as reflected by evaluation standards such as ITU-T P.800 \citep{itu1996telephone}, whereas the most rigorous intelligibility evaluations are multimodal, in that they require audio to be transcribed to text (cf.\ \citet{king2014measuring}).
Speech-synthesis literature also provides good support for the meaningfulness of separating the two aspects:
signal quality can be high even if speech is unintelligible, as demonstrated by the unconditional, ``babbling'' WaveNet system in \citet{oord2016wavenet}, whereas rule-based speech synthesisers can achieve ceiling intelligibility even if their naturalness is poor \citep{winters2004perception,malisz2019modern}.
A review of ten years of text-to-speech challenges found that different technical approaches on average performed better on different measures (conditional vs.\ unconditional) and worse on others \citep{king2014measuring}, even though these trends were not obvious from individual years due to the large variation among challenge submissions.

More generally, the roles of conditional and unconditional performance measures, and their interplay with each other, was formalised in a domain-agnostic context by \citet{blau2018perception} as the \emph{perception-distortion trade-off}.
\end{revision}

% Seems useful for a journal
\subsubsection{Motion aspects deliberately not evaluated}
Although an interesting question for a multispeaker dataset, we did not attempt to evaluate the appropriateness/specificity of the gesture motion style to different individuals in the database, since the data is too imbalanced to allow such an evaluation.
%Other facets of appropriateness, such as emotional appropriateness, or the appropriateness of motion for a specific individual speaker, were not evaluated.
Additionally, even though the speech and motion in the challenge comes from joint full-body motion capture of dyadic interactions with separate close-talking microphones for each speaker, the challenge only considered generating one side of the conversation, without awareness of the other party in the interaction (neither for the synthesis, \rev{nor} for the evaluation), in order to reduce problem complexity.

\subsection{Stimuli}
\subsubsection{Speech-segment selection}
The test set was deliberately made large to make it difficult to overfit to specific speech being evaluated.
Like the GENEA Challenge 2020 and
the Blizzard Challenges, not all test-set motion was included in the subjective evaluation.
From the 40 test-set chunks we selected 48 short \emph{segments} of test speech and corresponding test motion to be used in the subjective evaluations, based on the following criteria:
\begin{enumerate}
\item Segments should be around 8 to 10 seconds long, and ideally not shorter than 6 seconds.
\item The character should only be speaking, not passively listening, in the segments.
(No turn-taking, but backchannels from the interlocutor were OK.)
\item Segments should not contain any parts where \citet{lee2019talking} had replaced the speech by silence for anonymisation.
%(this appeared to affect mostly proper nouns).
\item Segments should be more or less complete phrases, starting at the start of a word and ending at the end of a word, and not end on a ``cliffhanger''.
\item \rev{The end of a segment should leave some margin until the chunk ends, to allow excerpting a longer segment if needed when creating mismatched stimuli as described in Sec.\ \ref{sssec:appropprocedure}.}
\item Finally, recorded motion capture in the segments (i.e., the FNA motion) should not contain any significant artefacts such as whole-body vibration or hands flicking open and closed due to poor finger tracking.
\end{enumerate}
The last item does not imply that the motion capture was perfect or completely natural for all segments in the evaluation, since the finger-tracking quality throughout the database does not allow our evaluations to reach that standard.
It merely means that the level of finger-tracking quality in the stimuli was consistent with the better parts of the source material from \citet{lee2019talking}.

The 48 segments we selected were between 5.6 and 12.1 seconds in duration (average 9.5 seconds).
Audio was loudness normalised to $-23$ dB LUFS following EBU R128 \cite{ebu2020loudness} to maintain a consistent listening volume in the user studies.
\begin{figure*}[!t]
\centering
%   \hfill
\begin{subfigure}[b]{0.416\textwidth}
    \includegraphics[width=\textwidth]{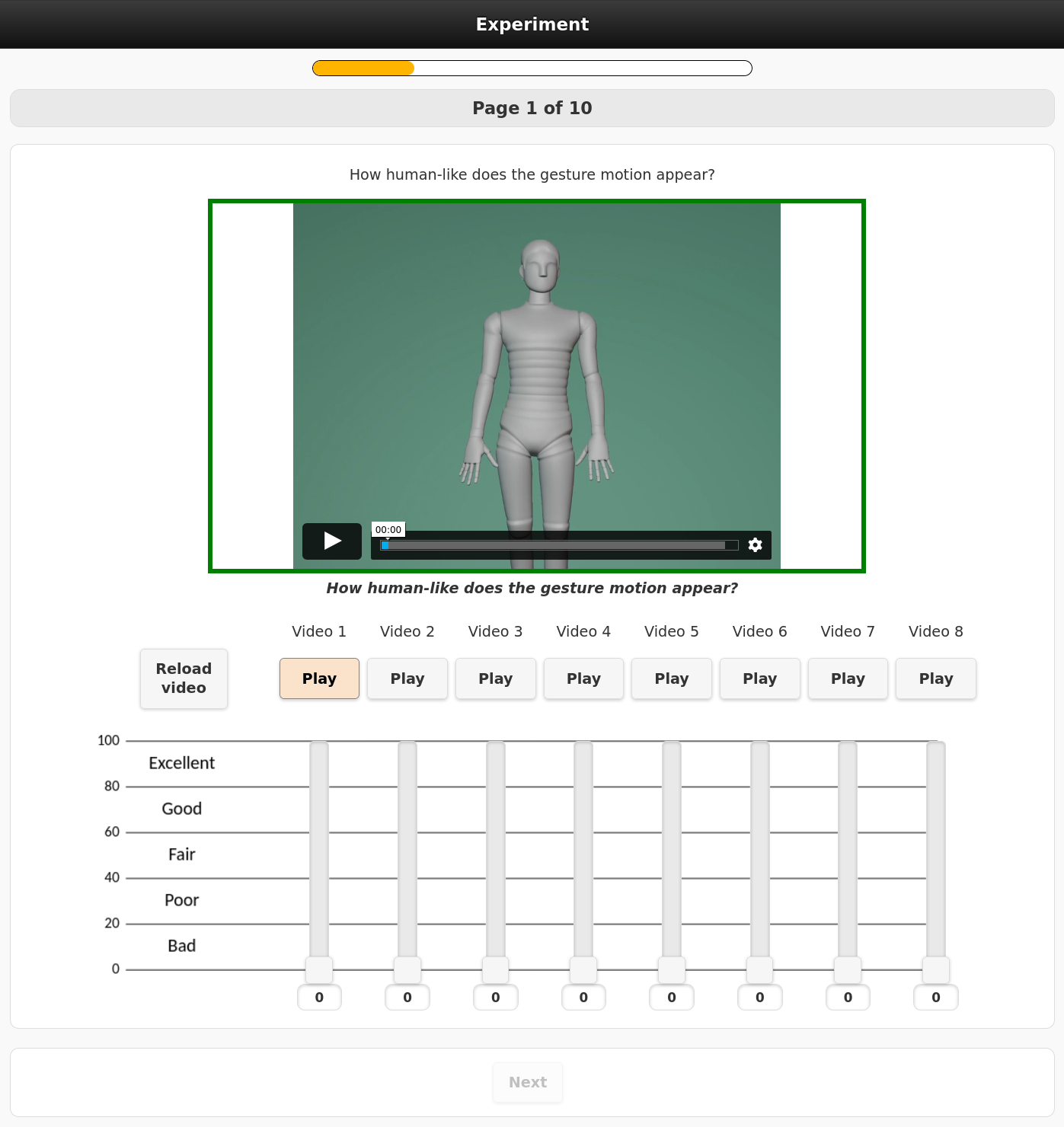}
    \caption{Human-likeness interface (HEMVIP) and full-body video}
    \label{fig:hemvipgui}
\end{subfigure}
\hfill\hfill
\begin{subfigure}[b]{0.564\textwidth}
    \centering\includegraphics[width=\columnwidth]{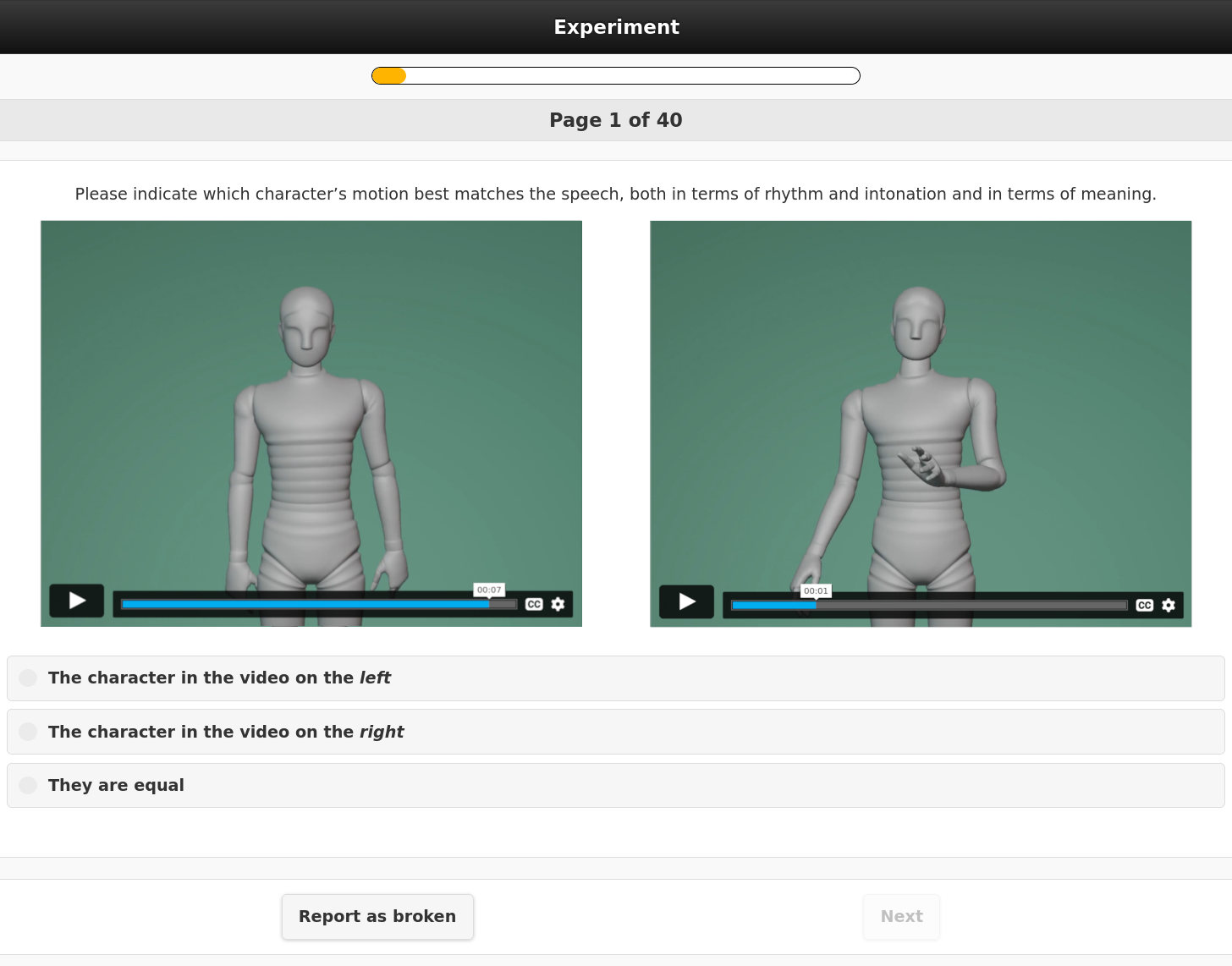}
    \caption{Appropriateness interface and upper-body videos}
    \label{fig:evaluation_interface_pairwise}
\end{subfigure}
%   \hfill
\vspace{-1.5pt}
\caption{Screenshots of the evaluation interfaces used in the studies, also showing the camera perspectives used by the \rev{two different} tiers.}
%\vspace{-2.6ex}
\label{fig:interfaces}
\Description{The left figure shows the human-likeness interface. The interface presents a video showing a full-body avatar and eight sliders for rating. There are play buttons for each slider and one reload video button. The right figure shows the appropriateness interface. The interface presents two side-by-side videos showing an upper-body avatar. There is a radio option group for the question asking which character's motion best matched the speech. The options are (1) the character in the video on the left, (2) the character in the video on the right, or (3) they are equal. A button marked Report as broken is on the bottom.}
\vspace{-1.5pt}
\end{figure*}

\subsubsection{Visualisation}
We used the same virtual avatar (shown in Fig.\ \ref{fig:interfaces}) in all rendered videos during the challenge and the evaluation.
\rev{The avatar had 56 joints (full body including fingers).
Since the speech and motion presented to our test takers was sourced from multiple people, the avatar was designed to be a gender-neutral humanoid figure without the hallmarks of any particular individual.
Instead of a fully realistic body shape and textures, a simplified design resembling a social robot was used.
As the challenge did not encompass the generation of gaze information, lip motion, or facial expression, eyes and mouth were omitted from the avatar, to help evaluators instead focus on the motion of the rest of the body.
%To reduce the risk of inducing the ``uncanny valley" effect on evaluators, the avatar does not adopt a fully realistic depiction of the human body, particularly in terms of textures and shape of body parts; the body looks rather "toony" or robotic-looking, and it does not use natural skin tones.
The avatar is of sufficiently high polygon count so that its surface was shaded smoothly.}
All teams had access to the official visualisation and rendering pipeline during the system-building phase, in the form of code, a portable Docker container, and a web server to which BVH files could be submitted to be rendered as video.
%\subsubsection{Visualisation server}
\label{subs:vis_ser}
%We also developed a visualisation server that enabled all participating teams to produce gesture-motion visualisations identical
%(except in resolution)
%to the video stimuli evaluated in the challenge.
%This was implemented using a headless Python-based web server which interfaced with \href{https://www.blender.org}{Blender} 2.93.9 and \href{https://ffmpeg.org/}{ffmpeg} to combine audio and video streams in a single video file.
Participants could send a 30-fps BVH file to the visualisation server, and these files would then be processed as quickly as possible into videos visualising the motion on the avatar.
%, in the order they came in (i.e., first in, first out).
%The server supports spawning multiple rendering workers, making it possible to do vertical scaling as long as the hardware resources allow it.
%The same visualisation was also used to render the final stimuli, in a resolution of 1440$\times$1080.
%but with the resolution increased to 1440$\times$1080 instead of the default 480$\times$270. The lower resolution was used to decrease the time spent rendering a video, consequently improving throughput of the server, since 30 teams initially took part in the challenge.
The visualisation server code has been open sourced (see \href{https://github.com/TeoNikolov/genea_visualizer}{github.com/TeoNikolov/genea\_visualizer}).
%and rendered stimulus videos are attached.
The code was available to the participants during the challenge, and they were free to use it to host their own servers if they wished.
The final rendered stimuli used a resolution of 1440$\times$1080.
%Videos are attached.

\subsection{Human-likeness evaluation}
\label{sec:humlike}

The human-likeness evaluation of the GENEA Challenge 2022 closely followed the human-likeness evaluation in the GENEA Challenge 2020 \citep{kucherenko2021large}.
%, by presenting \rev{groups of unimodal} motion \rev{video stimuli} and asking the subject to provide a rating for each \rev{video}.
\rev{Specifically,} the evaluation was based on the HEMVIP (Human Evaluation of Multiple Videos in Parallel) methodology \citep{jonell2021hemvip}, \rev{in which} multiple motion examples \rev{(video stimuli)} are presented \rev{on the same page (a.k.a.\ \emph{screen})} and the subject is asked to provide a rating for each one \rev{before continuing}.
All stimulus videos on the same page correspond to the same speech segment but different conditions.
\rev{This property of the HEMVIP method brings two advantages, namely (1) that any given study participant is always rating sets of stimuli for which the speech is the same and only the condition differs, which should make the numerical ratings and the relative condition ordering more consistent, and (2)} differences in rating between the different conditions can be analysed using pairwise statistical tests, which helps control for variation between different subjects and different input speech segments (as seen in the results in Sec.\ \ref{ssec:pairwisehumlike}).
%The videos used in this evaluation had the audio removed, since it has been found that speech and gesture perception influence each other \cite{bosker2021beat} and can confound motion evaluations \cite{jonell2020let}.
%(see Section\ \ref{ssec:pairwisehumlike}), where each observation was gathered from the same person watching videos featuring the same speech segment -- the only difference between videos on the same page being which condition that generated the motion of the virtual character in each video clip.
For a detailed explanation of the evaluation interface we refer the reader to \citet{jonell2021hemvip}, which introduced and validated the evaluation paradigm for gesture-motion stimuli.
Code is provided at \href{https://github.com/jonepatr/hemvip/tree/genea2022/}{github.com/jonepatr/hemvip/tree/genea2022/}.

\subsubsection{Evaluation design}
Each evaluation page asked participants ``How human-like does the gesture motion appear?'' and presented eight video stimuli to be rated on a scale from 0 (worst) to 100 (best) by adjusting an individual GUI slider for each video.
An example of the evaluation interface can be seen in Fig.\ \ref{fig:hemvipgui}. \rev{Note that by design only one video is visible at any given time; each play button corresponds to a distinct video stimulus, which is displayed when that button is clicked.}
Like in \citet{kucherenko2021large,jonell2021hemvip}, the 100-point rating scale was anchored by dividing it into successive 20-point intervals with labels (from best to worst) ``Excellent'', ``Good'', ``Fair'', ``Poor'', and ``Bad''.
These labels were based on those associated with the 5-point scale \rev{described} in
the Mean Opinion Score \rev{ITU-T P.800} standard \cite{itu1996telephone} for audio quality evaluation.

\begin{revision}
\subsubsection{Motivation for the use of unimodal stimuli}
\label{sssec:speechmotioninteraction}
Although the videos on any given page in these human-likeness evaluations all corresponded to the same speech input and had the same length, the videos presented to participants were unimodal (motion-only), in that they were completely silent and did not include any audio.
%, nor did the evaluation include any text transcriptions of the speech.
This ensures that ratings can only depend on the motion seen in the videos, and not on motion appropriateness for the speech, since raters did not have any access to any speech information.

Minimising the influence of speech is important when rating the intrinsic human-likeness of gesture motion, since speech and gesture perception are linked, to the extent that the same motion (possessing the same intrinsic human-likeness) may be perceived differently depending on what audio it is presented with.
A classic example of this link is the McGurk effect \citep{mcgurk1976hearing}, where the perceived identity of speech phonemes in a fixed audio stimulus changes if paired with video of specific facial articulation.
This effect also extends to gesture perception \citep{bosker2021beat}.
Conscious and unconscious human biases furthermore mean that raters may give lower or higher scores based on their perception of speaker traits such as likeability, gender, social status, etc.\ (e.g., \citet{babel2015expectations,montgomery2018intergroup}), which would increase estimator variance and reduce statistical resolution.
Removing speech content and only presenting the motion on a neutral avatar avoids these issues.
More explicitly, \citet{jonell2020let} found that including speech audio in a study of motion mimicry in annotated reference stimuli led to confounding, since subjects based their responses on speech semantic content rather than the relevant non-verbal interactions in the video modality.
Consequently, our evaluation is unimodal and the task given to the raters is not conditional on the speech.
%Since it has been found that speech and gesture perception influence each other \cite{bosker2021beat} and can confound motion evaluations \cite{jonell2020let},
\end{revision}

\subsubsection{Evaluation procedure}
The test was preceded by a screen with instructions, which the participants would read.
Then, each subject completed one training page showing a fixed set of videos with different motion, to familiarise participants with the task and what the stimuli would look like, before starting the study in earnest.
The training phase was followed by 10 pages of ratings for the evaluation.
Responses given on the training page were not included in the analysis.
The evaluation was balanced such that each segment appeared on pages 1 through 10 with approximately equal frequency across all participants (segment order), and each condition was associated with each slider with approximately equal frequency across all pages (condition order).
For any given participant and study, each of the 10 pages would use a different speech segment.
Every page in the evaluation contained one stimulus video from condition FNA/UNA. %, visualising the motion-capture data gathered from the speaker at the same time that they uttered the speech in question.
This was used to help calibrate evaluators' ratings and keep them consistent throughout the test.
Since motion-capture data projected onto a virtual character may not necessarily be perceived as
perfectly natural, there was no requirement to rate the best motion as 100.
After completing the rating pages, but before submitting the study, participants filled in a short questionnaire to gather broad, anonymous demographic information, the results of which are presented in Sec.\ \ref{subs:att_checks_hl}.

\subsection{Appropriateness evaluation}
\label{sec:approp}
The appropriateness evaluation was designed to assess the link between the \rev{gesture} motion and the input speech, separate from the intrinsic human-likeness of the motion.
\rev{It is thus inexorably multimodal, with user assessments of motion conditional on speech information provided to them.}

In the previous GENEA Challenge, appropriateness was evaluated using a HEMVIP-based rating study very similar to that for human-likeness, except that speech audio was included in the videos.
\rev{In an attempt to control for the effect of motion quality in that evaluation,} test takers were asked to ignore the motion quality and only rate the appropriateness of the motion for the speech \cite{kucherenko2021large}.
Unfortunately, that evaluation was not altogether successful, since their \emph{mismatched} condition M -- which paired natural motion segments with unrelated speech segments, intended as a bottom line -- attained the second-highest appropriateness rating, above all synthetic systems.
This suggests a significant \rev{interaction} between the perceived human-likeness of a motion segment and its perceived appropriateness for speech.
\rev{That interaction} acted as a \rev{confounder} in their study, with the result that all submissions ranked below natural-looking motion unrelated to the speech, \rev{despite the latter being} intended as a bottom line in terms of appropriateness.

\subsubsection{Evaluation design}
For the GENEA Challenge 2022, we decided to evaluate motion appropriateness for speech in a different way.
Our design goal for \rev{this} challenge was to assess appropriateness whilst controlling for the human-likeness of the motion in an effective way.
To do so, we took the idea of motion mismatching like in \citet{jonell2020let} \rev{(which studied facial motion rather than hand and body gestures)} and used it within every condition\rev{, and} not just for the recorded motion-capture data FNA/UNA\rev{.}

On each page, subjects were presented with a pair of videos containing the same speech audio.
Both videos contained motion from the same \rev{source -- i.e., the same condition --} and \rev{were thus expected to have similar} motion quality \rev{and motion characteristics (at least on average)}, but one was matched to the speech audio and the other mismatched, belonging to unrelated speech.
Whether the left or the right video was mismatched was randomised.
Subjects were then asked to ``Please indicate which character's motion best matches the speech, both in terms of rhythm and intonation and in terms of meaning.''
In response, they could choose the character on the left, on the right, or indicate that the two were equally well matched (``They are equal'', also referred to as \emph{equal}\ or a \emph{tie}).
We asked for preferences rather than ratings since there is evidence \cite{wolfert2021rate} that this is more efficient in pairwise comparisons like these.
A screenshot of the evaluation interface used for the appropriateness studies is presented in Fig.\ \ref{fig:evaluation_interface_pairwise}.

The extent to which test-takers prefer the character with the matched motion reveals how specific the gesture motion is to the given speech: random motion will result in a 50--50 split, whereas conditions whose motion is more specifically appropriate to the input speech are expected to elicit a higher relative preference for the matched motion.
In this type of evaluation, condition M (the mismatched condition) from the 2020 challenge will perform at chance rate, rather than being tied for second highest as in 2020.

\citet{rebol2021passing} used a similar methodology with preference tests to quantify the correlation (essentially, the appropriateness) between generated hand and body gestures and their associated speech, which we were not aware of until after conducting our challenge.
However, they asked a different question of the users, did not quantify the appropriateness of real human motion, \rev{used data from monocular video rather than 3D motion capture (leading to noticeably lower data quality),} and only used the approach to evaluate a single gesture-generation method.

\rev{Since speech audio has to be present in our appropriateness evaluation stimuli, test-taker perception may be subject to the human biases discussed in Sec.\ \ref{sssec:speechmotioninteraction}.
However, the fact that the speech information (and the avatar used) is the exact same on both sides in every pairwise video presentation in the preference test should serve to control for the effect of these biases on user responses.}

\subsubsection{Evaluation procedure}
\label{sssec:appropprocedure}
Concretely, we created the mismatched stimuli by taking the 48 existing speech and motion segments \rev{selected for our evaluations}, and \rev{permuting} the motion in between them such that no motion segment ever remained in its original place.
\rev{Each motion segment thus featured twice in the pairwise study: once with matched speech, and once with mismatched speech, where the corresponding matched stimulus would use another motion segment from the same source condition.}
%Mathematicians call such a permutation a \emph{derangement}.
As the 48 different segments did not all have the same length, a longer or shorter segment of motion generally had to be excerpted from the motion chunks (original or generated), so as to match the new speech duration.
\rev{This was done by moving the endpoint of the mismatched motion segment such that the resulting motion duration exactly matched that of the new speech audio.}
The starting point of the \rev{mismatched} motion video was \rev{never changed, and was thus} always the same as in the respective matched stimulus video (i.e., corresponding to the start of a phrase).
%The mismatched video stimuli are available at \href{}{here}.

After an instruction page and a training page, each subject evaluated 40 pages with one pair of videos each.
This means that subjects watched 80 videos total in each study, the same number of videos as was evaluated in the human-likeness studies (ignoring the training pages in all cases).
Each study was balanced such that each speech segment, condition, and order of the two videos appeared approximately equally many times.
\begin{table*}
\caption{Demographics of test takers from the four user studies. Gender ``F'' stands for ``female'', ``M'' for ``male'', and ``X'' for ``prefer not not say''.}
\label{tab:demographics}
\centering
\small%
\begin{tabular}{@{}ll|c|cccccc|ccc|c@{}}
\toprule
 &  & Total number & \multicolumn{6}{c|}{Country of residence} & \multicolumn{3}{c|}{Gender} & Age (years)\tabularnewline
Study & Tier & of test takers & AUS & CAN & IE & NZ & UK & USA & F & M & X & mean $\pm$ std.~dev.\tabularnewline
\midrule
Human-likeness & Full body & 121 & 2 & \hphantom{0}2 & \hphantom{0}3 & 0 & 110 & \hphantom{0}4 & \hphantom{0}60 & \hphantom{0}60 & 1 & 38 $\pm$ 12\tabularnewline
 & Upper body & 150 & 1 & \hphantom{0}0 & \hphantom{0}4 & 0 & 134 & 11 & \hphantom{0}74 & \hphantom{0}75 & 1 & 40 $\pm$ 13\tabularnewline
Appropriateness & Full body & 247 & 3 & 13 & 10 & 2 & 211 & \hphantom{0}8 & 137 & 107 & 3 & 38 $\pm$ 14\tabularnewline
 & Upper body & 304 & 2 & 10 & \hphantom{0}1 & 0 & 256 & 35 & 127 & 173 & 4 & 38 $\pm$ 13\tabularnewline
\bottomrule
\end{tabular}
\end{table*}

\subsection{Test takers and attention checks}
\label{subs:att_checks_hl}
It has recently been found that crowdsourced evaluations are not significantly different from in-lab evaluations in terms of results and consistency \cite{jonell2020iva_crowd}.
The challenge therefore adopted an entirely crowdsourced approach, as opposed to, for example, the Blizzard Challenge, which has used a mixed approach. Attention checks were used to exclude participants that were not paying attention.
%, as detailed in Section\ \ref{subs:att_checks_hl}.
Test takers (a.k.a.\ subjects) were recruited through the crowdsourcing platform \href{https://www.prolific.co/}{Prolific}.
We used Prolific's built-in pre-screening tools to restrict the pool of test-takers in two ways: (1) subjects were required to reside in any of six English-speaking countries, namely Australia, Canada, Ireland, New Zealand, the United Kingdom, and the USA, and (2) subjects were required to have English as their first language.

We conducted four user studies, two for human-likeness and two for appropriateness.
A subject could take one or more studies, but could only participate in each study at most once, and could not use a phone or tablet to take the test.

Each study incorporated four attention checks per person, to make sure that subjects were paying attention to the task and remove insincere test-takers.
For the human-likeness studies, these attention checks took the form of a text message ``Attention! You must rate this video NN'' superimposed on the video.
``NN'' would be a number from 5 to 95, and the subject had to set the corresponding slider to the requested value, plus or minus 3, to pass that attention check.
Which sliders on which pages that were used for attention checks was uniformly random, except that no page had more than one attention check, and the natural motion (condition FNA and UNA) was never replaced by an attention check.
For the appropriateness studies, the attention checks either displayed a brief text message over the gesticulating character, reading ``Attention! Please report this video as broken'', or they temporarily replaced the audio with a synthetic voice speaking the same message.
Subjects were exposed to two attention checks of each kind.
To pass the attention check, participants had to click a button marked ``Report as broken'' seen in Fig.\ \ref{fig:evaluation_interface_pairwise}, forwarding them to the next pair of videos in the evaluation.
Since reporting a video as broken avoids having to give a response, it can in theory be used to quickly skip through the test.
To help prevent this, we implemented the button such that it becomes clickable after a 5-second delay, after the page is loaded.
However, as this does not fully prevent skipping through the test, subjects who used that button more than three times on pages without attention checks were also removed without pay.
In all studies, the attention-check messages did not appear until a few seconds into each attention-check video, so that participants who only watched the first seconds would be unlikely to pass the checks.

Subjects who failed two or more attention checks were removed from the respective study without being paid, since Prolific's policies do not allow rejecting a subject on the basis of a single failed attention check.
Only the subjects who failed zero or one attention check for a study have been included in our analyses below.
Responses to videos used for attention checks were not included in our analyses.
Right before submitting their results, subjects also filled in a short questionnaire to gather broad, anonymous demographic information about the population taking the test.

A design goal of the human-likeness studies was that every combination of two distinct conditions should appear on the pages approximately equally often, and at least 600 times (not counting FNA/UNA, which appeared on every page).
To meet this goal, we recruited 121 test takers that successfully passed the attention checks and completed the full-body study, and 150 test takers that
%successfully passed the attention checks and completed
did the same for the upper-body study.
Of these, all passed all attention checks, except for one subject in the upper-body study, who failed one attention check.
Since the upper-body study compared 11 conditions instead of only 10, it required more raters to reach the desired number of ratings pairs.
Table \ref{tab:demographics} provides demographic details of all subjects in the user studies.
%Of the 121 test takers in the full-body study, 60 identified as female, 60 as male, and 1 did not want to disclose their gender.
%The same numbers for the 150 upper-body test takers were 74, 75, and 1, respectively.
%For the full-body test takers, 2 resided in Australia, 2 in Canada, 3 in Ireland, 110 in the United Kingdom, and 4 in the USA.
%The mean age for the full-body study was 37.89 with a standard deviation of 11.47. 
%The upper-body study had 1 participant residing in Australia, 4 in Ireland, 134 in the United Kingdom, and 11 in the USA.
%The mean age for the upper-body study was 39.68 with a standard deviation of 12.59. 
%The mean age of the test takers was 38 years with a standard deviation of 12 in the full-body study and 40 years with a standard deviation of 13 in the upper-body study.

For the appropriateness studies, our design goal was for each condition to receive as many responses per condition as the number of ratings that each condition (aside from FNA/UNA) received in the corresponding human-likeness evaluation.
This works out to 880 responses per condition in the full-body studies and 990 responses per condition in the upper-body studies.
Because a subject in these studies provided half as many responses as in a human-likeness study (40 vs.\ 80), the appropriateness studies needed to recruit approximately twice as many test takers.
In the end, 247 test takers successfully passed the attention checks in the full-body study, while 304 passed the attention checks in the upper-body study.
All of these passed all attention checks, except for 10 participants in the full-body study and 14 participants in the upper-body study, who each failed one attention check.
Demographic details are provided in Table \ref{tab:demographics}.
%10 participants failed 1 attention check in the full-body study, and 14 missed 1 attention check in the upper-body study. 
%
%Of the 247 participants in the full-body study, 137 identified as female, 107 as male, and 3 did not want to disclose their gender.
% full body study: The mean age was 38 with standard deviation 14).
% upper body study: The mean age was 38 with standard deviation 13).
%The same numbers for the 304 upper-body test takers were 127, 173, and 4, respectively.
%For the full-body test takers, 3 resided in Australia, 13 in Canada, 10 in Ireland, 2 in New Zealand, 211 in the United Kingdom, and 8 in the USA.
%The upper-body study had 2 participants residing in Australia, 10 in Canada, 1 in Ireland, 256 in the United Kingdom, and 35 in the USA.
%The mean age of the test takers was 38 years in both studies, with a standard deviation of 14 for the full-body study and 13 for the upper-body study.
%All of these passed all attention checks, except for 10 participants in the full-body study and 14 participants in the upper-body study, who each failed one attention check.
Each subject in a study contributed 36 ratings to the analyses after removing attention checks, unless they had to skip a page in the rare case of a video failing to load (which occurred approximately 1.6 times per 1000 videos presented).

Test takers were remunerated 6 GBP for each successfully completed human-likeness study.
Since the median completion time was 28 minutes each, this corresponds to a median compensation just above 12 GBP per hour.
Similarly, the appropriateness studies took a median of 24 or 25 minutes to complete, and earned a reward of 5.5 GBP each, amounting to around 13 GBP per hour.
These compensation levels all exceed the UK national living wage and also exceeds the highest living wage quoted by the Living Wage Foundation in the UK at the time of the evaluation.
All numbers are as measured by Prolific, which uses the median rather than the mean for these calculations to prevent extreme completion times from skewing the data.
%``so that extreme completion times do not skew the data''.
%
Response data from the evaluation, together with statistical analysis code, is provided at \href{https://doi.org/10.5281/zenodo.6939888}{doi.org/10.5281/zenodo.6939888}.

\subsection{Objective metrics}
\label{ssec:obj_metrics}
The main goal of the GENEA challenge is to compare human subjective impressions of the outputs of different gesture-generation systems.
\rev{We therefore discourage} using the results of automated performance metrics as indicators of the perceptual impressions of different systems.
%achieve human-like and appropriate speech-driven automatic gesture generation, we discourage using the results of automated performance metrics as indicative the merits of a system.
However, since subjective evaluation is costly and time-consuming, it would be beneficial for the field to identify meaningful objective evaluation methods, especially for use during system development.
%In addition to the subjective evaluation, we also computed a number of different objective metrics on the conditions in the challenge evaluation.
As a step in this direction we therefore considered five objective measures previously used to evaluate co-speech gestures, namely average jerk, average acceleration, distance between gesture speed (i.e., absolute velocity) histograms, canonical correlation analysis, and the Fréchet distance between motion feature distributions.
We computed these metrics for each condition in each tier using the complete test sequences, i.e., not only on the motion segments featured in the subjective evaluation.
Details on each metric are provided below. The code for the numerical evaluations has been made publicly available to enhance reproducibility.\footnote{See \href{https://github.com/genea-workshop/genea_numerical_evaluations}{github.com/genea-workshop/genea\_numerical\_evaluations}.}

To compare and validate these metrics against our subjective evaluation, we provide results on the rank correlations between subjective and objective metrics in Sec.\ \ref{ssec:objectiveresults}. 

\subsubsection{Average acceleration and jerk}
The third time derivative of the joint positions is called \emph{jerk} and can be defined mathematically as
$\textrm{jerk}(\boldsymbol{x}) = \boldsymbol{x}'''(t)$.
%\begin{align}
%\mathrm{jerk}(\boldsymbol{x}) = \boldsymbol{x}'''(t)
%\text{.}
%\end{align}
The average value of the absolute magnitude of the jerk is commonly used to quantify motion smoothness \citep{morasso1981spatial,uno1989formation,kucherenko2019analyzing}.
We report average values of absolute jerk (defined using finite differences) averaged across all test motion segments. A perfectly natural system should have average jerk very similar to natural motion.

We also evaluated the same measure, but computed using the absolute value of the acceleration
%\begin{align}
%\mathrm{acc.}(\boldsymbol{x}) = %\boldsymbol{x}''(t)
%\text{,}
%\end{align}
$\textrm{acc.}(x) = x''(t)$
instead of the jerk.
Again, we expect natural-looking motion to have similar average acceleration as in the reference data.

\subsubsection{Comparing speed histograms}
The distance between speed histograms has also been used to evaluate gesture quality \citep{kucherenko2019analyzing,kucherenko2020gesticulator}, since well-trained models should produce motion with similar properties to that of the actor it was trained on.
In particular, it should have a similar motion-speed profile for any given joint.
This metric is based on the assumption that synthesised motion should follow a speed distribution \rev{similar to the motion-capture data}.
To evaluate this similarity we calculate speed-distribution histograms for all systems and compare them to the speed distribution of natural motion (condition N) by computing the Hellinger distance \citep{nikulin2001hellinger},
\begin{align}
H(\boldsymbol{h}^{(1)},\, \boldsymbol{h}^{(2)})
& = \sqrt{1 - \sum_{i}{\sqrt{h^{(1)}_i \cdot h^{(2)}_i}}}
\text{,}
\end{align}
between the histograms $\boldsymbol{h}^{(1)}$ and $\boldsymbol{h}^{(2)}$.
Lower distance is better.

For both of the objective evaluations above the motion was first converted from joint angles to 3D coordinates. 

\subsubsection{Canonical correlation analysis}

Canonical correlation analysis (CCA) \citep{thompson1984canonical} is a form of linear subspace analysis, and involves the projection of two sets of vectors
(here the generated poses and those from FNA/UNA)
%(here, the variable sets $\boldsymbol{x}$ and $\boldsymbol{y}$)
onto a joint subspace.
CCA has been used to evaluate gesture-generation models in previous work \citep{sadoughi2019speech, bozkurt2015affect,jinhong_lu_2020_4088376}.

The goal of CCA is to find a sequence of linear transformations of each variable set, such that the Pearson correlation between the transformed variables is maximised. This correlation is what we use as a similarity measure, and we report it as global CCA values in our results.
A high value is considered better.
\begin{table*}[!t]
\centering%
\caption{Summary statistics of responses from all user studies, with 95\% confidence intervals. ``M.'' stands for ``matched'' and ``Mism.'' for ``mismatched''. ``Percent matched'' identifies how often subjects preferred matched over mismatched motion. \rev{Human-likeness values are between 0 and 100; higher is better.}}
\label{tab:stats}
%\hfill
\begin{subtable}[t]{0.49\textwidth}
\centering%
\caption{Full-body}%
\label{stab:fbstats}%
\small%
\begin{tabular}{@{}l|c|ccc|c@{}}
\toprule
& Median
& \multicolumn{4}{c@{}}{Appropriateness} \\
& human- & \multicolumn{3}{c|}{Num.\ responses}
& Percent matched\\
ID & likeness%Median rating
& M. & Tie & Mism. & (splitting ties)\\
\midrule
FNA & $70\hphantom{.0}\in{}[69,71]$
& 590 & 138 & 163 & $74.0\in{}[70.9,76.9]$\\
FBT & $27.5\in{}[25,30]$
& 278 & 362 & 250 & $51.6\in{}[48.2,55.0]$\\
FSA & $71\hphantom{.0}\in{}[70,73]$
& 393 & 216 & 269 & $57.1\in{}[53.7,60.4]$\\
FSB & $30\hphantom{.0}\in{}[28,31]$
& 397 & 163 & 330 & $53.8\in{}[50.4,57.1]$\\
FSC & $53\hphantom{.0}\in{}[51,55]$
& 347 & 237 & 295 & $53.0\in{}[49.5,56.3]$\\
FSD & $34\hphantom{.0}\in{}[32,36]$
& 329 & 256 & 302 & $51.5\in{}[48.1,54.9]$\\
FSF & $38\hphantom{.0}\in{}[35,40]$
& 388 & 130 & 359 & $51.7\in{}[48.2,55.1]$\\
FSG & $38\hphantom{.0}\in{}[35,40]$
& 406 & 184 & 319 & $54.8\in{}[51.4,58.1]$\\
FSH & $36\hphantom{.0}\in{}[33,38]$
& 445 & 166 & 262 & $60.5\in{}[57.1,63.8]$\\
FSI & $46\hphantom{.0}\in{}[45,48]$
& 403 & 178 & 312 & $55.1\in{}[51.7,58.4]$\\
\bottomrule
\end{tabular}
\end{subtable}%
\hfill\hfill
\begin{subtable}[t]{0.49\textwidth}
\centering%
\caption{Upper-body}%
\label{stab:ubstats}%
\small%
\begin{tabular}{@{}l|c|ccc|c@{}}
\toprule
& Median
& \multicolumn{4}{c@{}}{Appropriateness} \\
& human- & \multicolumn{3}{c|}{Num.\ responses}
& Percent matched\\
ID & likeness%Median rating
& M. & Tie & Mism. & (splitting ties)\\
\midrule
UNA & $63\hphantom{.0}\in{}[61,65]$
& 691 & 107 & 189 & $75.4\in{}[72.5,78.1]$\\
UBA & $33\hphantom{.0}\in{}[31,34]$
& 424 & 264 & 303 & $56.1\in{}[52.9,59.3]$\\
UBT & $36\hphantom{.0}\in{}[34,39]$
& 341 & 367 & 287 & $52.7\in{}[49.5,55.9]$\\
USJ & $53\hphantom{.0}\in{}[52,55]$
& 461 & 164 & 365 & $54.8\in{}[51.6,58.0]$\\
USK & $41\hphantom{.0}\in{}[40,44]$
& 454 & 185 & 353 & $55.1\in{}[51.9,58.3]$\\
USL & $22\hphantom{.0}\in{}[20,25]$
& 282 & 548 & 159 & $56.2\in{}[53.0,59.4]$\\
USM & $41\hphantom{.0}\in{}[40,42]$
& 503 & 175 & 328 & $58.7\in{}[55.5,61.8]$\\
USN & $44\hphantom{.0}\in{}[41,45]$
& 443 & 190 & 352 & $54.6\in{}[51.4,57.8]$\\
USO & $48\hphantom{.0}\in{}[47,50]$
& 439 & 209 & 335 & $55.3\in{}[52.1,58.5]$\\
USP & $29.5\in{}[28,31]$
& 440 & 180 & 376 & $53.2\in{}[50.0,56.4]$\\
USQ & $69\hphantom{.0}\in{}[68,70]$
& 504 & 182 & 310 & $59.7\in{}[56.6,62.9]$\\
\bottomrule
\end{tabular}
\end{subtable}%
%hfill
\end{table*}

\subsubsection{Fréchet gesture distance} 
Recent work by \citet{yoon2020speech} proposed the Fréchet gesture distance (FGD) to quantify the quality of generated gestures. 
This metric is based on the FID metric used in image-generation studies \citep{heusel2017gans} and can be written
%
%%% START OF COMMENT BLOCK - May be useful for future reference
%We follow the FGD setup proposed by Yoon et al. \citet{yoon2020speech}, and train an auto-encoder as a feature extractor on the Human3.6M dataset from \citet{ionescu2013human3}. A convolutional encoder and decoder form the basis of our feature extractor, the encoder is trained to encode a set of direction vectors $d$ to a latent feature $z^{gesture}$. The decoder is set up to decode the latent features into a set of direction vectors, reconstructing the pose sequence the encoder was given as input. Like in \citet{yoon2020speech}, FGD($X,\hat{X}$) is defined as the Fréchet distance between the Gaussian mean and the covariance of the latent features of the ground-truth gestures $X$ and the Gaussian mean and the covariance of the latent features of the generated gestures $\hat{X}$:
%\centering
%%% END OF COMMENT BLOCK
\begin{align}
\mathrm{FGD}(\boldsymbol{X},\,\hat{\boldsymbol{X}})
& = ||\boldsymbol{\mu}_r - \boldsymbol{\mu}_g||^2 + \mathrm{tr}(\boldsymbol{\Sigma}_r + \boldsymbol{\Sigma}_g - 2(\boldsymbol{\Sigma}_r\boldsymbol{\Sigma}_g)^{1/2})
\text{.}
\end{align}
Here, $\boldsymbol{\mu}_r$ and $\boldsymbol{\Sigma}_r$ are the first and second moments of the latent-feature distribution $\boldsymbol{Z}_r$ of the human motion-capture data $\boldsymbol{X}$, whereas $\boldsymbol{\mu}_g$ and $\boldsymbol{\Sigma}_g$ are the first and second moments of the latent-feature distribution $\boldsymbol{Z}_g$ of the generated gestures $\hat{\boldsymbol{X}}$. $\boldsymbol{Z}_r$ and $\boldsymbol{Z}_g$ were extracted by the same feature extractor, which was obtained as the encoder part of a motion-reconstructing autoencoder.
We used a CNN-based autoencoder trained on the challenge dataset following the implementation in \citet{yoon2020speech}.
Lower values are better.

\subsubsection{System ranking comparison}
A good objective metric might help in evaluating the performance of a system, especially when such a metric correlates with a subjective measure.
To get more insight into whether the objective metrics in our study may be used as a proxy for subjective evaluation results, we calculated the correlation between the ranking of the conditions on median human-likeness, and the result on the objective metrics listed above.
For this, we used Kendall's $\tau$ rank correlation coefficient, and associated statistical tests \cite{kendall1970rank}. 

Of the objective metrics we studied, only CCA compares output poses directly to the corresponding reference motion-capture poses.
All other metrics are invariant to permutation, in the sense that changing the order of the different sequences (mismatching them with other speech/reference motion) will not change the value.
%of mismatching the generated motion with other speech features and reference motion.
They thus cannot measure appropriateness, which is why we only consider how those metrics correlate with human-likeness scores.
%we only compute rank correlations between these metrics and the human-likeness scores.

\section{Results}
\label{sec:results}
\begin{revision}
The results of the challenge are significant and thought provoking. It is the first time that we find a system generating 3D gesture motion that exceeds the source data in terms of human-likeness, whilst simultaneously laying bare the extent of the gap between natural and synthetic gesture motion in terms of their appropriateness for speech. We furthermore find that all objective metrics except for the FGD correlate so poorly with subjective human-likeness scores as to be statistically indistinguishable from chance correlation.
More detail is provided in the sections below, first reporting the results of the subjective evaluation and thereafter the objective metrics.
Discussion of the various findings is reserved for Sec.\ \ref{sec:discussion}.
\end{revision}

%The results of the challenge are revolution and a revelation, for the first time finding performance that exceeds the ground-truth data in human-likeness, whilst simultaneously laying bare the true extent of the gap between natural and synthetic motion in terms of speech appropriateness.
%We furthermore find that all objective metrics except for the FGD correlate so poorly with subjective scores as to be statistically indistinguishable from chance correlation.
%More detail is provided in the sections below, first reporting the results of the subjective evaluation and thereafter the objective metrics.
%Discussion of the various findings is reserved for Sec.\ \ref{sec:discussion}.

\subsection{Analysis and results of human-likeness studies}
\label{ssec:pairwisehumlike}
\begin{figure*}[t!]
\centering%
%\hfill
  \begin{subfigure}[b]{0.49\textwidth}
    \centering%
    \includegraphics[width=\textwidth]{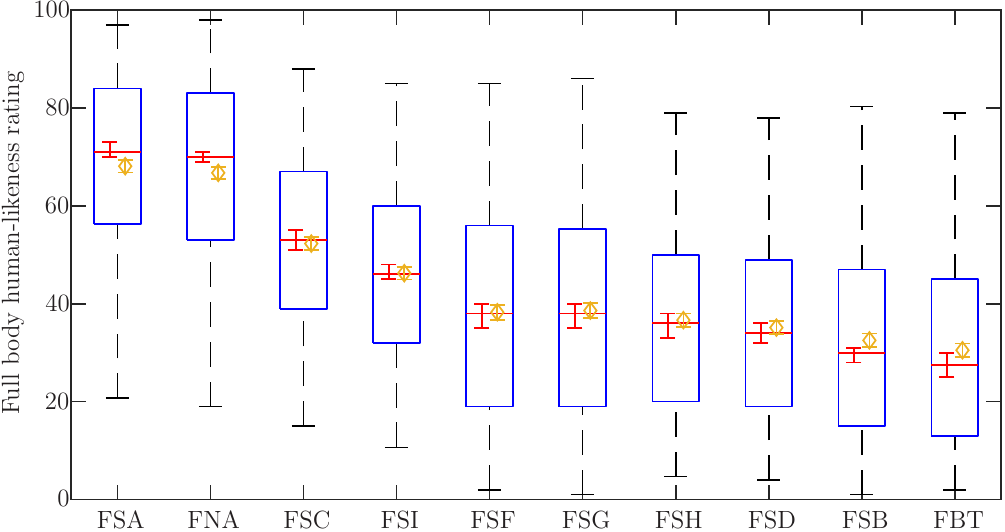}
    \caption{Full-body}
    \label{sfig:fbhumlikeboxplot}
  \end{subfigure}
\hfill
  \begin{subfigure}[b]{0.49\textwidth}
    \centering%
    \includegraphics[width=\textwidth]{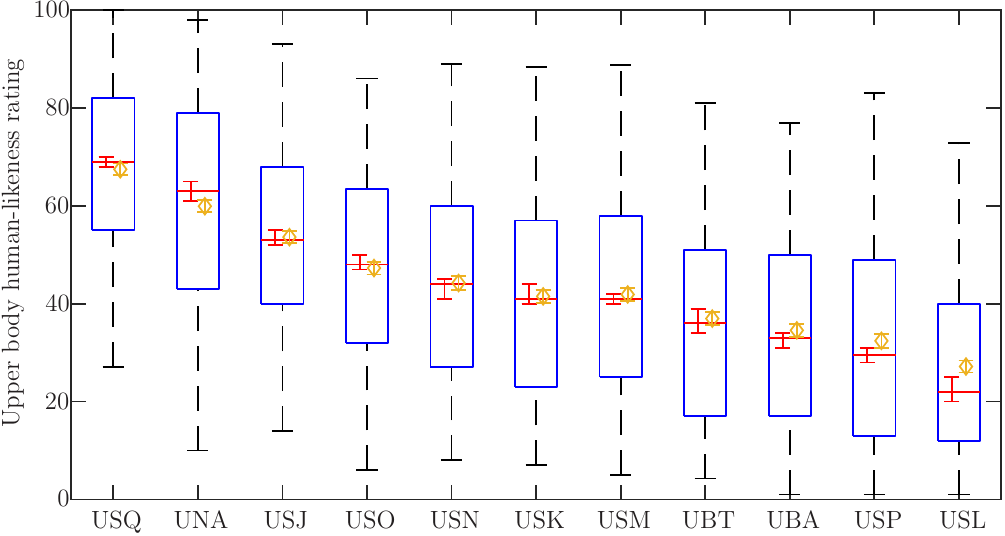}
    \caption{Upper-body}
    \label{sfig:ubhumlikeboxplot}
  \end{subfigure}
%\hfill
\vspace{-2.65pt}
\caption{Box plots visualising the ratings distribution in the human-likeness studies. Red bars are medians and yellow diamonds are means, each with a 0.05 confidence interval and a Gaussian assumption for the means. Box edges are at 25 and 75 percentiles, while whiskers cover 95\% of all ratings for each condition. Conditions are ordered descending by sample median for each tier.}
\label{fig:humlikeboxplots}
\Description{The first box plot shows full-body human-likeness ratings for the 10 conditions. The conditions are sorted in descending order, based on their rating; the order is FSA, FNA, FSC, FSI, FSF, FSG, FSH, FSD, FSB, and FBT. FSA shows a median rating of 71. FBT shows a median rating of 27.5. The second box plot shows upper-body human-likeness ratings for the 11 conditions. The conditions are sorted in descending order, based on their rating; the order is USQ, UNA, USJ, USO, USN, USK, USM, UBT, UBA, USP, and USL. USQ shows a median rating of 69. USL shows a median rating of 22.}
\vspace{-2.65pt}
\end{figure*}

Each test taker in the human-likeness studies contributed 76 ratings to the analyses after removing attention checks, giving a total of 9,196 ratings for the full-body study and 11,400 ratings for the upper-body study.
The results are visualised in Fig.\ \ref{fig:humlikeboxplots},
with summary statistics (sample median and sample mean) for the ratings of all conditions in each of the two human-likeness studies given in the first half of Table\ \ref{tab:stats}, together with 95\% confidence intervals for the true median.
These confidence intervals were computed using order statistics, leveraging the binomial distribution cdf, while those for the mean used a Gaussian assumption (i.e., using Student's $t$-distribution cdf, rounded outward to ensure sufficient coverage); see \citet{hahn1991statistical}.
We note that statistics regarding the mean should be interpreted with caution, since responses should be seen as ordinal rather than numerical, and it is therefore improper from a perceptual perspective to perform averaging on the ratings.
\begin{figure*}[t!]
\centering%
%\hfill
  \begin{subfigure}[b]{0.49\textwidth}
    \centering%
    \includegraphics[width=\textwidth]{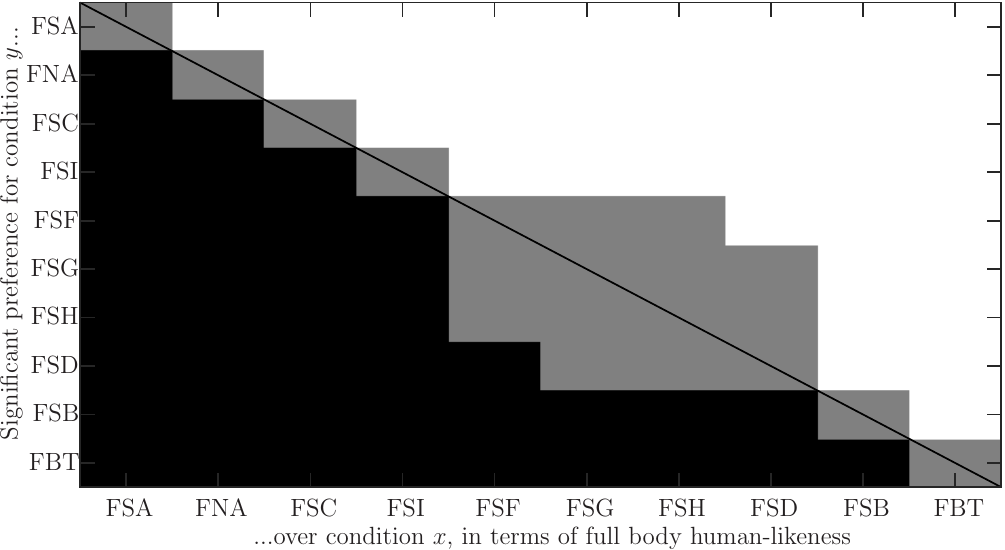}
    \caption{Full-body}
    \label{sfig:fbhumlikedifferences}
  \end{subfigure}
%\hfill
  \begin{subfigure}[b]{0.49\textwidth}
    \centering%
    \includegraphics[width=\textwidth]{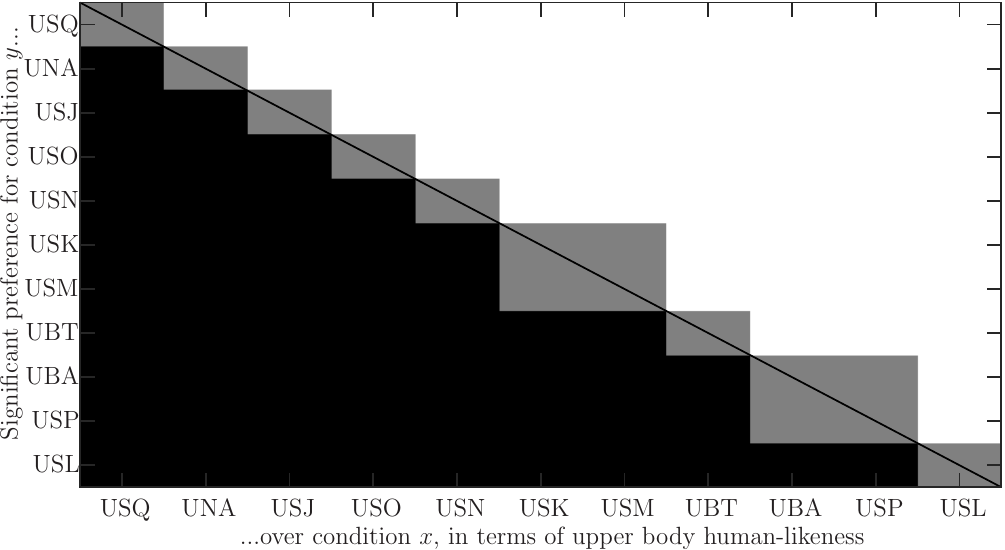}
    \caption{Upper-body}
    \label{sfig:ubhumlikedifferences}
  \end{subfigure}
%\hfill
\vspace{-2.65pt}
\caption{Significant differences in human-likeness. White means the condition listed on the $y$-axis rated significantly above the condition on the $x$-axis, black means the opposite ($y$ rated below $x$), and grey means no statistically significant difference at level $\alpha=0.05$ after Holm-Bonferroni correction. Conditions use the same order as the corresponding subfigure in Figure\ \ref{fig:humlikeboxplots}.}
\label{fig:humlikedifferences}
\Description{These two plots visualise significant differences between every two condition pairs. The first plot shows the results of the full-body human-likeness study. There are 10 conditions, namely: FSA, FNA, FSC, FSI, FSF, FSG, FSH, FSD, FSB, and FBT. FSA, FNA, FSC, FSI, FSB, and FBT showed significantdifferences over all the other conditions. FSF did not show significant preferences over FSG and FSH. FSG did not show significant preferences over FSF, FSH, and FSD. FSH did not show significant preferences over FSF, FSG, and FSD. FSD did not show significant preferences over FSG and FSH. The second plot shows the results of the upper-body human-likeness study. There are 11 conditions: USQ, UNA, USJ, USO, USN, USK, USM, UBT, UBA, USP, and USL. All the conditions except USK, USM, UBA, and USP showed significant differences over all the other conditions. USK did not show significant preferences over USM. USM did not show significant preferences over USK. UBA did not show significant preferences over USP. USP did not show significant preferences over UBA.}
\vspace{-2.65pt}
\end{figure*}

The distributions in Fig.\ \ref{fig:humlikeboxplots} are seen to be quite broad.
This is common in evaluations like HEMVIP \cite{jonell2021hemvip}, since the range of the responses not only reflects differences between conditions, but also extraneous variation, e.g., between stimuli, in individual preferences, and in how critical different raters are in their judgements.
In contrast, the plotted confidence intervals are seen to be quite narrow, since the statistical analysis can mitigate the effects of much of this variation.

Despite the wide range of the distributions, the fact that the conditions were rated in parallel on each page enables using pairwise statistical tests to factor out many of the above sources of variation.
To analyse the significance of differences in median rating between different conditions, we applied two-sided pairwise Wilcoxon signed-rank tests to all unordered pairs of distinct conditions in each study.
(This is the same methodology as in the GENEA Challenge 2020 \cite{kucherenko2021large}.)
This closely follows the analysis methodology used throughout recent Blizzard Challenges and, unlike Student's $t$-test (which assumes that rating differences follow a Gaussian distribution), this analysis is valid also for ordinal response scales, like those we have here.
For each condition pair, only cases where both conditions appeared on the same page and were assigned valid ratings were included in the analysis of significant differences.
(Recall that not all conditions were rated on all pages due to the limited number of sliders and the presence of attention checks.)
This meant that every statistical significance test was based on at least 615 pairs of valid ratings in the full-body study, and 603 pairs of valid ratings in the upper-body study.
Because this analysis is based on pairwise statistical tests, it can potentially resolve differences between conditions that are smaller than the width of the confidence intervals for the median in Fig.\ \ref{fig:humlikeboxplots}, since those confidence intervals are inflated by variation that the statistical test controls for.
The $p$-values computed in the significance tests were adjusted for multiple comparisons on a per-study basis using the Holm-Bonferroni method \cite{holm1979simple}, which is uniformly more powerful than conventional Bonferroni correction at keeping the family-wise error rate (FWER), often referred to as alpha-level, at or below $\alpha=0.05$

Our statistical analysis found all but 5 out of 45 condition pairs to be significantly different in the full-body study and all but 2 out of 55 condition pairs to be significantly different in the upper-body study, all at the level $\alpha=0.05$ after Holm-Bonferroni correction.
The significant differences we identified in the two studies are visualised in Fig.\ \ref{fig:humlikedifferences} which uses the same condition order as the box plot and shows which conditions were found to be rated significantly above or below which other conditions.

\subsection{Analysis and results of appropriateness studies}
We gathered a total of 8,867 responses for the full-body study and 10,910 responses from the upper-body study that were included in the analysis.
Every condition received at least 873 responses in the full-body study and 983 in the upper-body study.
\begin{figure*}[t!]
\centering%
%\hfill
  \begin{subfigure}[b]{0.49\textwidth}
    \centering%
    \includegraphics[width=\textwidth]{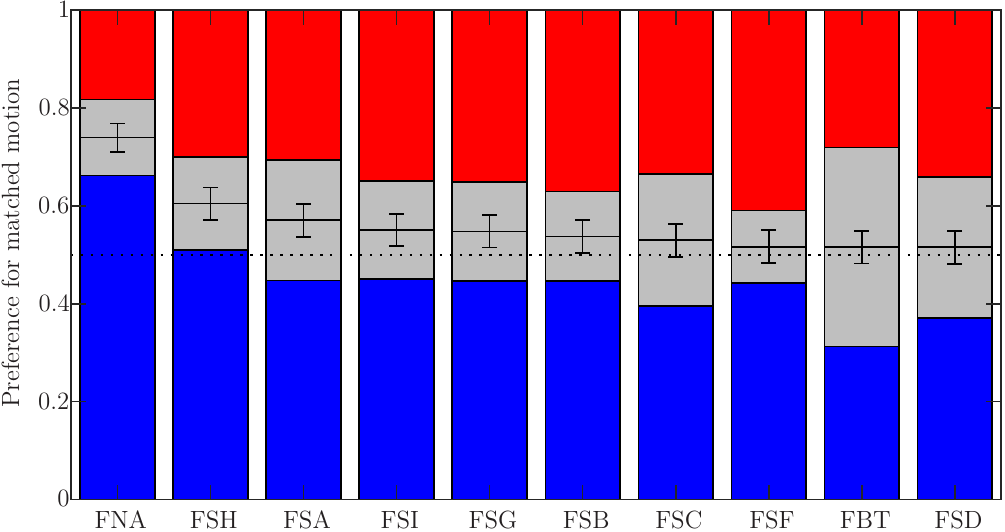}
    \caption{Full-body}
    \label{sfig:fbappropbars}
  \end{subfigure}
\hfill
  \begin{subfigure}[b]{0.49\textwidth}
    \centering%
    \includegraphics[width=\textwidth]{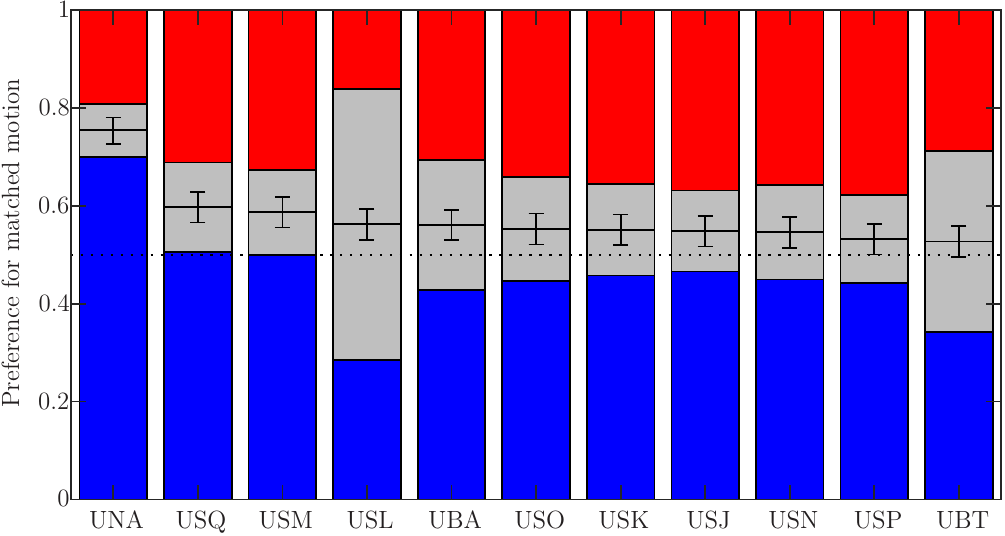}
    \caption{Upper-body}
    \label{sfig:ubappropbars}
  \end{subfigure}
%\hfill
\caption{Bar plots visualising the response distribution in the appropriateness studies. The blue bar (bottom) represents responses where subjects preferred the matched motion, the light grey bar (middle) represents tied (``They are equal'') responses, and the red bar (top) represents responses preferring mismatched motion, with the height of each bar being proportional to the fraction of responses in each category. The black horizontal lines bisecting the light grey bars represent the proportion of matched responses after splitting ties, each with a 0.05 confidence interval. The dotted black line indicates chance-level performance. Conditions are ordered by descending preference for matched motion after splitting ties.}
\label{fig:appropbars}
\Description{The first stacked bar chart shows the preference of matched versus mismatched full-body motion for 10 conditions. The conditions are sorted in descending order, based on the preference for the matching condition. The order is FNA, FSH, FSA, FSI, FSG, FSB, FSC, FSF, FBT, and FSD. For FNA, the blue, gray, and red regions take about 65 percent, 15 percent, and 20 percent, respectively. The second tacked bar chart shows the preference of matched versus mismatched upper-body motion for 11 conditions. The descending order of preference for the matching motion is UNA, USQ, USM, USL, UBA, USO, USK, USJ, USN, USP, and UBT. For UNA, the blue, gray, and red regions take about 70 percent, 10 percent, and 20 percent, respectively.}
\end{figure*}
\begin{figure*}[t!]
\centering%
%\hfill
  \begin{subfigure}[b]{0.49\textwidth}
    \centering%
    \includegraphics[width=\textwidth]{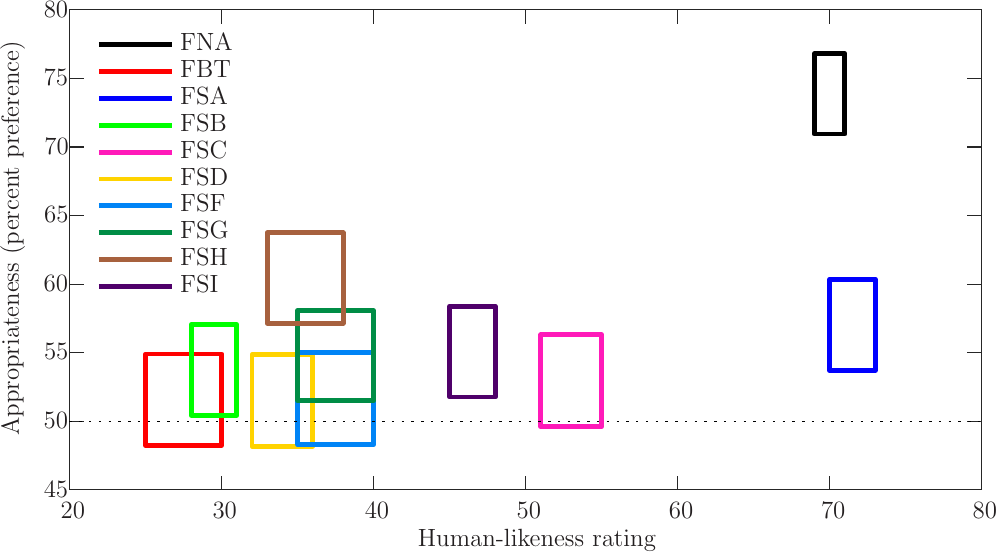}
    \caption{Full-body}
    \label{sfig:fbjpoint}
  \end{subfigure}
\hfill
  \begin{subfigure}[b]{0.49\textwidth}
    \centering%
    \includegraphics[width=\textwidth]{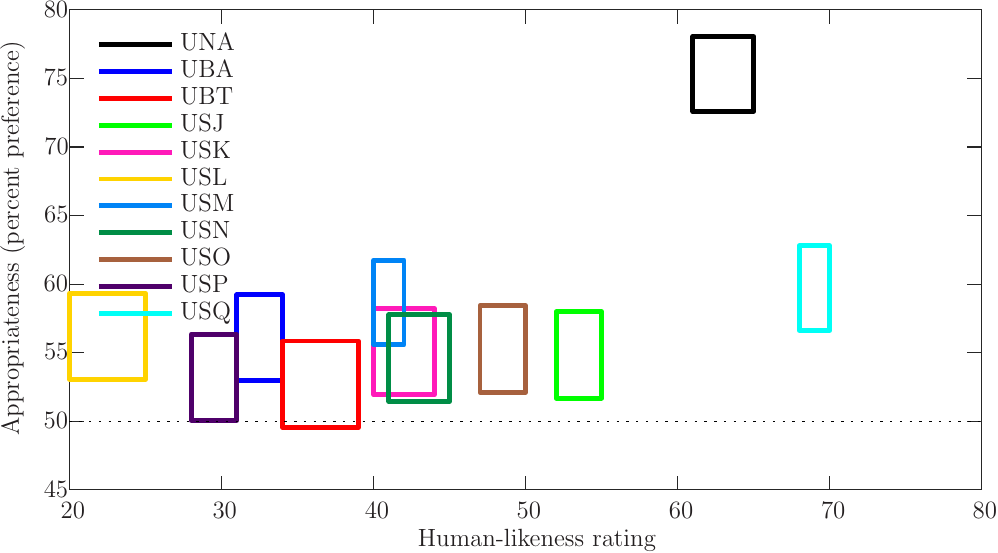}
    \caption{Upper-body}
    \label{sfig:ubjoint}
  \end{subfigure}
%\hfill
\caption{Joint visualisation of the evaluation results for each tier. Box widths show 95\% confidence intervals for the median human-likeness rating and box heights show 95\% confidence intervals for the preference for matched motion in percent, indicating appropriateness.}
\label{fig:joint}
\Description{There are two plots, one for the full-body tier and one for the upper-body tier. In both plots, human-likeness is on the x-axis, and appropriateness (in percentages) on the y-axis.  In the first plot for full-body, the FNA box is in the top right corner, which means high ratings on human-likeness and appropriateness. The FSA box is on the right side. FSC and FSI are on the middle bottom. The other boxes are in the bottom left corner. In the upper-body plot, the UNA box is in the top right corner. The USQ box is on the right side (further right than UNA). The other boxes are in the bottom left corner (around 55 on the y-axis, spread between 20 and 55 on the x-axis).}
\end{figure*}
Raw response statistics for all conditions in each of the two studies are shown in the second half of Table\ \ref{tab:stats}, together with 95\% Clopper-Pearson confidence intervals for the fraction of time that the matched video was preferred over the mismatched, after dividing ties equally between the two groups (rounding up in case of non-integer counts).
The confidence intervals were rounded outward to ensure sufficient coverage.
The response distributions in the two studies are further visualised through bar plots in Fig.\ \ref{fig:appropbars}, whilst Fig.\ \ref{fig:joint} visualises the results of the entire challenge in a single coordinate system per tier.

Unlike the human-likeness studies, the responses in the appropriateness studies are restricted to three categories and do not necessarily come in pairs for statistical testing in the same way as for the parallel sliders in HEMVIP.
A different method for identifying significant differences therefore needs to be adopted.
We used Barnard's test \cite{barnard1945new} to identify statistically significant differences at the level $\alpha=0.05$ between all pairs of distinct conditions, applying the Holm-Bonferroni method \cite{holm1979simple} to correct for multiple comparisons as before.
(Here and forthwith, we only consider the relative preference in the sample after dividing ties equally.)
Barnard's test is considered more appropriate than Fisher's exact test for a product of two independent binomial distributions \cite{lydersen2009recommended}, as here.

Our statistical analysis found 13 of 45 condition pairs to be significantly different in the full-body study and 10 out of 55 condition pairs to be significantly different in the upper-body study.
Specifically, FNA/UNA were significantly more appropriate for the specific speech signal compared to all other, synthetic conditions.
In addition, FSH was significantly more appropriate than FBT, FSC, FSD, and FSF in the full-body study.
As before, the significant differences we identified in the two studies are visualised in Fig.\ \ref{fig:humlikedifferences} which uses the same condition order as the box plot and shows which conditions were found to be rated significantly above or below which other conditions.
No other pairwise differences were statistically significant in either study.
\begin{table*}[!t]
\centering%
\caption{Objective evaluation results.
The word ``acceleration'' has been abbreviated to ``accel.'';
$\pm$ shows the standard deviation per sequence.
The best two or three numbers in each column, i.e., those closest to the numbers from the held-out motion-capture data (FNA/UNA, first row of values), are bold.
Except for FNA/UNA, conditions (rows) are ordered by decreasing median human-likeness rating.
%Condition labels in green indicates one of the best-performing systems in terms of human likeness, while red colour marks the worst-performing systems interms to the human-likeness user study) for easier comparison.
Numbers have generally been rounded to three significant digits.
%Note no strong relation between subjective and objective measures.
}
\label{tab:obj_stats}
%\hfill
%\hfill
\begin{subtable}[t]{0.49\textwidth}
\centering%
\caption{Full-body}%
\label{stab:fb_obj}%
\small%
\begin{tabular}{@{}l|ccccc@{}}
\toprule
 & Average & Average & Global & Hellinger & FGD \\
Condition & jerk & accel. & CCA & distance & \\
\midrule
                              FNA   &     \tablebf{31300 $\pm$ 6590} & \tablebf{798 $\pm$ 208} & \tablebf{1\hphantom{.000}}    & \tablebf{0\hphantom{.000}}     & \tablebf{\hphantom{0}0\hphantom{.00}}\\
\midrule
{%\color[HTML]{38761D}
{FSA}} &            14600 $\pm$ 2970  & \tablebf{668 $\pm$ 161} &         0.849  & \tablebf{0.041} &             \tablebf{3.18}           \\
{%\color[HTML]{38761D}
{FSC}} & \hphantom{0}5130 $\pm$ 2120  &         332 $\pm$ 129  &         0.818  &         0.125  &                    16.4\hphantom{0} \\
{%\color[HTML]{38761D}
{FSI}} & \hphantom{0}7370 $\pm$ 1710  &         345 $\pm$ \hphantom{0}98   &         0.789  &         0.111  & \tablebf{\hphantom{0}4.87}           \\
                              FSF   &    \tablebf{22600 $\pm$ 6240} & \tablebf{666 $\pm$ 223} & \tablebf{0.916} &         0.195  &         \hphantom{0}7.49            \\
                              FSG   & \hphantom{0}5560 $\pm$ 2380  &         282 $\pm$ 127  & \tablebf{0.992} & \tablebf{0.060} &                    10.1\hphantom{0} \\
                              FSH   & \hphantom{0}8630 $\pm$ 2440  &         313 $\pm$ \hphantom{0}92   & \tablebf{0.968} &         0.104  & \tablebf{\hphantom{0}4.02}           \\
                              FSD   & \hphantom{0}8690 $\pm$ 8320  &         405 $\pm$ 257  &         0.886  &         0.132  &                    43.4\hphantom{0} \\
{%\color[HTML]{990000}
{FSB}} &    \tablebf{27200 $\pm$ 4680} & \tablebf{628 $\pm$ 116} &         0.782  & \tablebf{0.050} &                    16.3\hphantom{0} \\
{%\color[HTML]{990000}
{FBT}} & \hphantom{0}3510 $\pm$ 1090  &         177 $\pm$ \hphantom{0}56   &         0.738  &         0.267  &                    28.6\hphantom{0} \\
%\noalign
% \cr
\bottomrule
% \cr
\end{tabular}
\end{subtable}%
\hfill\hfill
%\vspace{8pt}
\begin{subtable}[t]{0.49\textwidth}
\centering%
\caption{Upper-body}%
\label{stab:up_obj}%
\small%
\begin{tabular}{@{}l|ccccc@{}}
\toprule
 & Average & Average & Global & Hellinger & FGD \\
Condition & jerk & accel. & CCA & distance & \\
\midrule
                               UNA   &   \tablebf{33000 $\pm$ \hphantom{0}7030} & \tablebf{842 $\pm$            222} & \tablebf{1\hphantom{.000}} & \tablebf{0\hphantom{.000}} & \tablebf{\hphantom{00}0\hphantom{.00}}         \\
\midrule
{%\color[HTML]{38761D}
{USQ}} &   \tablebf{15400 $\pm$ \hphantom{0}3190} & \tablebf{710 $\pm$            173} &                    0.685   &           \tablebf{0.043}  &           \tablebf{\hphantom{00}2.84}          \\
{%\color[HTML]{38761D}
{USJ}} & \hphantom{0}8280 $\pm$ \hphantom{0}1460  &          375 $\pm$ \hphantom{0}81  &                    0.640   &                    0.197   &           \tablebf{\hphantom{00}4.83}          \\
{%\color[HTML]{38761D}
{USO}} & \hphantom{0}5450 $\pm$ \hphantom{0}2260  &          353 $\pm$            138  &                    0.812   &                    0.129   &                    \hphantom{0}16.4\hphantom{0}\\
                               USN   & \hphantom{0}7510 $\pm$ \hphantom{0}3400  &          384 $\pm$            127  &                    0.789   &                    0.092   &                               194\hphantom{.00}\\
                               USK   & \hphantom{0}8180 $\pm$ \hphantom{0}2450  &          311 $\pm$ \hphantom{0}99  &           \tablebf{0.962}  &                    0.137   &                    \hphantom{0}15.5\hphantom{0}\\
                               USM   & \hphantom{0}6840 $\pm$ \hphantom{0}3200  &          385 $\pm$            172  &           \tablebf{0.991}  &           \tablebf{0.039}  &           \tablebf{\hphantom{00}2.17}          \\
                               UBT   & \hphantom{0}3760 $\pm$ \hphantom{0}1170  &          190 $\pm$ \hphantom{0}60  &                    0.707   &                    0.248   &                    \hphantom{0}18.2\hphantom{0}\\
                               UBA   &   \tablebf{18000 $\pm$            14900} &          513 $\pm$            326  &           \tablebf{0.964}  &                    0.244   &                    \hphantom{0}17.0\hphantom{0}\\
{%\color[HTML]{990000}
{USP}} &   \tablebf{28500 $\pm$ \hphantom{0}4960} & \tablebf{661 $\pm$            123} &                    0.769   &           \tablebf{0.051}  &                    \hphantom{0}18.0\hphantom{0}\\
{%\color[HTML]{990000}
{USL}} & \hphantom{0}7730 $\pm$ \hphantom{0}5420  &          258 $\pm$            157  &                    0.849   &                    0.306   &                    \hphantom{0}28.4\hphantom{0}\\
\bottomrule
\end{tabular}
\end{subtable}%
%\vspace{-1\baselineskip}
\end{table*}%
\begin{table*}[!t]
\caption{Rank correlations (Kendall's $\tau$) between the ``error'' in the objective metrics (how much each objective value differed from the reference FNA/UNA) and median human-likeness scores (here abbreviated ``Hum.'') or -- only for CCA -- the preference for matched motion after splitting ties (abbreviated ``App.'').
A strong predictor of human scores will exhibit a $\tau$-value close to negative unity combined with a low $p$-value.}
    \begin{subtable}[t]{\columnwidth}
        \centering%
        \caption{Full-body}
        \small%
        \begin{tabular}{@{}l|cccccc@{}}
        \toprule
        Metric & Average & Average & \multicolumn{2}{c}{Global} & Hellinger & FGD \\
        & jerk & accel. & \multicolumn{2}{c}{CCA} & distance & \\
        Versus & Hum. & Hum. & Hum. & App. & Hum. & Hum. \\
            \midrule
            $\tau$ & $-0.09$ & $-0.36$ & $-0.36$ & $-0.38$ & $-0.36$ & $-0.49\hphantom{0}$ \\
            $p$-value & $\hphantom{-}0.72$ & $\hphantom{-}0.15$ & $\hphantom{-}0.16$ & $\hphantom{-}0.15$ & $\hphantom{-}0.15$ & $\hphantom{-}0.048$ \\
            \bottomrule
        \end{tabular}
    \end{subtable} 
\hfill\hfill
    \begin{subtable}[t]{\columnwidth}
        \centering%
        \caption{Upper-body}
        \small%
        \begin{tabular}{@{}l|cccccc@{}}
         \toprule
         Metric & Average & Average & \multicolumn{2}{c}{Global} & Hellinger & FGD \\
         & jerk & accel. & \multicolumn{2}{c}{CCA} & distance & \\
            Versus & Hum. & Hum. & Hum. & App. & Hum. & Hum. \\
            \midrule
            $\tau$ & $-0.11$ & $-0.26$ & $0.11$ & $-0.49\hphantom{0}$ & $-0.40\hphantom{0}$ & $-0.51\hphantom{0}$ \\
            $p$-value & $\hphantom{-}0.64$ & $\hphantom{-}0.27$ & $0.64$ & $\hphantom{-}0.041$ & $\hphantom{-}0.085$ & $\hphantom{-}0.029$ \\
            \bottomrule
        \end{tabular}
    \end{subtable}
    %\hfill\hfill
\label{tab:rankcorr}
%\vspace{-1\baselineskip}
\end{table*}

Instead of comparing the appropriateness of different synthesis approaches against one another, one may instead compare to a random baseline (50/50 performance), and test if the observed effect size is statistically significantly different from zero.
We can assess this at the 0.05 level by checking whether or not the confidence interval on the effect size overlaps with chance performance.
From this perspective, \rev{FNA,} FSA, FSB, FSG, FSH, FSI are significantly more appropriate than chance in the full-body study, and all \rev{conditions} except UBT are more appropriate than chance in the upper-body study.
Unlike other significance tests in the subjective evaluation, these assessments do not include a correction for multiple comparisons.

\subsection{User comments}
As part of the post-evaluation questionnaire, we asked study participants to comment on the user studies, including positive and negative aspects they perceived.
97\% of the respondents in the user studies responded positively on whether the compensation was adequate.
Additionally, they often commented positively on how interesting and engaging the study was.

We also asked participants regarding any negative aspects of the study.
Here, 15\% of the participants answered that they found repetitiveness a negative aspect of the study.
Some users pointed at the lack of a proper human face on the humanoid, and suggested incorporating that in future work.
Others commented on the lack of real conversation, and proposed to have the humanoid be part of an actual conversation.
All responses to these questions can be found in our data release.

\subsection{Objective evaluation results}
\label{ssec:objectiveresults}
\begin{figure*}[!t]
\centering
\includegraphics[trim={3.2cm 1.4cm 3.2cm 1.4cm},clip,width=\textwidth]{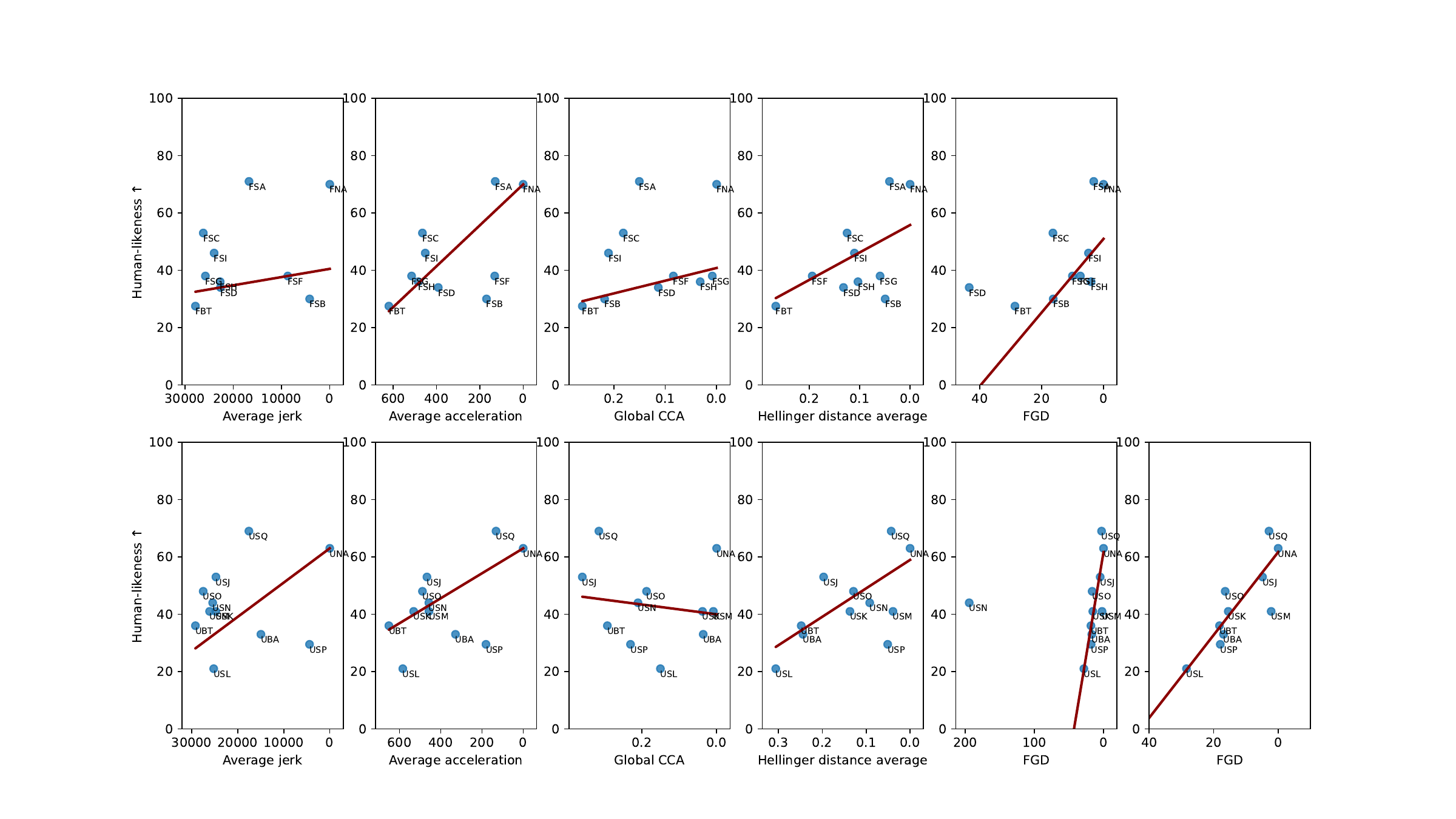}
\caption{Scatterplots comparing objective metrics and human-likeness ratings. The first row is for the full-body tier and the second row is for the upper-body tier.
The $x$-axis shows the absolute magnitude of the difference between the objective value for each system and the corresponding value for the reference motion FNA/UNA, with the scale reversed such that the systems most similar to the reference are on the right.
%The $x$-axis represents the absolute value of the difference between each value and human motion's value for each metric (lower is better; human motion FNA and UNA are always zero and they are on the right end because the $x$-axis is reversed).
Regression lines (from the Theil-Sen regressor \cite{Theil1992,sen1968estimates}, which is robust to outliers) are also shown. The last plot in the second row is for FGD but with a narrower $x$-axis range for a better view.}
\label{fig:objectivemetric}
\Description{
There are 11 scatterplots, 5 on the first row and 6 on the second. The plots depict comparisons between objective metrics and human-likeness ratings across the two tiers: full-body on the first row, and upper-body on the second row. The x-axis shows the absolute magnitude of the difference between the objective value for each system and the corresponding value for the reference motion FNA/UNA, with the scale reversed such that the systems most similar to the reference are on the right. The y-axis shows the human-likeness rating between 0 and 100. All evaluated systems are plotted onto each scatterplot and marked by a dot. A regression line is also drawn. Each scatterplot addresses one objective metric, in order, from the following list of objective metrics: Average jerk, Average acceleration, Global canonical correlation analysis, Hellinger distance average, and FGD. The 6th scatterplot on the second row is another view of the FGD metric, but with a narrower x-axis range for a better view; this addresses the compressed view of all but one metrics in the 5th scatterplot on row 2, where an outlier system USN is present.}
\vspace{-0.5\baselineskip}
\end{figure*}

The values of the objective metrics we computed are listed in Table\ \ref{tab:obj_stats}. 
For each number in the table, we also calculated how much it differed from the corresponding value for the reference system (FNA/UNA), and then computed the rank correlation between the absolute value of these differences and the median human-likeness scores from the subjective evaluation.
The idea is that systems exhibiting values closer to FNA/UNA should appear more human-like.
The resulting rank correlations and $p$-values can be found in Table \ref{tab:rankcorr}.
For median human-likeness, we only found a statistically significant ($p<0.05$) rank correlation with FGD, for both the full and upper-body tier (Kendall's $\tau=-0.49$ and $-0.51$, respectively).
The negative sign is expected, since a smaller difference from FNA/UNA should be associated with better-looking motion and higher human-likeness scores.
Fig.\ \ref{fig:objectivemetric} visually compares the subjective human-likeness ratings and objective metric results.

CCA is the only metric we computed that can indicate appropriateness, since it directly compares each generated sequence to the corresponding reference motion-capture poses.
We therefore computed its rank correlations with the appropriateness data as well.
Here we found a statistically significant effect ($\tau=-0.49$) for the upper-body tier, but not for the full body.

\section{Discussion}
\label{sec:discussion}
We now discuss our results and how they may be interpreted, first for human-likeness (in Sec.\ \ref{ssec:humlikecomments}), then for appropriateness (in Sec.\ \ref{ssec:appropriatenesscomments}), and then for the objective metrics (in Sec.\ \ref{ssec:obj_discussion}).
We connect our discussion of each part to the other evaluations we performed and to previous literature.
Based on our findings, we then formulate a number of take-home messages regarding what matters most in gesture generation (in Sec.\ \ref{ssec:whatlearnt}) and give examples of how materials from the challenge can be used by the field (in Sec.\ \ref{ssec:materials}).

\subsection{Discussion of human-likeness results}
\label{ssec:humlikecomments}
Generating \rev{convincing} human-like gestures is a difficult problem, and nearly all conditions rated significantly below natural motion capture.
However, each tier contains an entry which is rated significantly above the motion from the motion-capture recordings in terms of human-likeness.
This is a leap forwards compared to GENEA 2020, and we believe it represents a motion quality not before seen in large-scale evaluations.
Although there has been work, specifically \citet{rebol2021passing}, that reported a proposed motion-generation method as being statistically not significantly different from natural motion, they only evaluated a single method and their study was not based on motion-capture data but on 3D pose estimation from monocular video.
We think that that choice of data source restricted the motion quality of their natural-motion condition to be less convincing (and thus \rev{easier to surpass}) than our reference-motion conditions FNA/UNA.
Furthermore, all differences between natural and synthetic conditions are significant in our study.

\subsubsection{Interpreting the high scores of FSA and USQ}
Despite \citet{zhou2022gesturemaster} (conditions FSA and USQ) being rated above the corresponding natural reference motion, we caution that this does not mean that the motion is ``superhuman'', or even completely human-like -- indeed, the median rating is much below 100, which would constitute ``completely human-like'' as per the instructions to test takers.
What the result means is rather that the visualised motion in the majority of cases was perceived as more human-like than the
%motion-capture in the database, specifically than the
motion-capture data used for FNA/UNA in the subjective evaluation.
In making this distinction, it is important to keep in mind that our human-likeness evaluation is constrained by several factors.
Most notably, the nominally natural motion is constrained by our ability to accurately capture the entire range of human motion, especially the fingers, using the technology we used.
Finger motion capture is very difficult, and dataset limitations meant that the finger motion could not be chosen so as to look completely natural in all test segments evaluated, potentially degrading the ratings of FNA/UNA as a result.
An artificial system might have its training data cleaned of problematic instances, so as to prevent it from generating such motion, giving it an edge over FNA/UNA.
This is in fact what was done for systems FSA and USQ, which only used selected training-data segments, manually chosen to have high motion quality, in generating their output gestures \cite{zhou2022gesturemaster}.

Our ability to visualise human characters and their motion also plays a role in our findings.
The use of a deliberately neutral 3D avatar lacking potentially distracting human features such as gaze and lip motion significantly reduces the bandwidth of the communication channel to the user, which lowers the threshold for what needs to be achieved in order to match human motion ratings in the evaluation.
If the challenge had involved generating additional modalities such as gaze and facial expression, the shortcomings of artificial systems may have become more clear, at the expense of increased complexity when running and taking part in the challenge.
%In addition, the greater interquartile range of ratings of UNA compared to FNA could mean that the process of imposing full-body motion from a walking and talking human onto an avatar with fixed lower body may not always yield completely natural results.
%Future GENEA Challenges intend to only consider full-body motion.

\subsubsection{On the differences between the two tiers}
There are fewer significant differences in the full-body evaluation than in the upper-body evaluation, perhaps meaning that full-body motion is more difficult to rate consistently.
Although the difference is not substantial, we would naively expect the opposite, due to the correction for multiple comparisons being more conservative in the upper-body evaluation.
There are many possible explanations for this finding, beyond the fact that the different teams did not all participate in both tiers.
For example, our finding is consistent with an interpretation that full-body motion is a more difficult machine-learning problem, for instance due to increased dimensionality of the output space and the increased number of behaviours that need to be learnt.
This could explain why the best entry in the upper-body evaluation more clearly outperformed UNA, compared to the margin between the best entry in the full-body evaluation and FNA.

Another possible explanation for the same result is that the process of imposing full-body motion from a walking and talking human onto an avatar with a fixed lower body may not always yield completely natural results, and could sometimes give rise to incongruous motion.
This could also explain the wider span (greater interquartile range) of ratings of UNA compared to FNA.
Future GENEA challenges intend to only consider full-body motion.

\subsection{Discussion of appropriateness results}
\label{ssec:appropriatenesscomments}
We find the results of the appropriateness evaluation both thought-provoking and revealing about the state of the field.
To begin with, the greatest relative preference, a 75\% preference for matched motion, was observed for natural motion capture, i.e., FNA/UNA.
This +25\% effect size over the 50/50 bottom line validates that our methodology can well identify when gestures are appropriate for the speech and is about half the theoretical maximum value of +50\% (a 100/0 split).
A +25\% effect size should be considered a good result, since previous studies that have incorporated mismatched stimuli, e.g., \citet{jonell2020let,rebol2021passing}, have found that they sometimes are difficult for participants to distinguish from matched ones, especially if they -- like here -- both correspond to segments where the character is speaking (and do not, say, match audio of active speaking with a segment of motion corresponding to the character listening without speaking; cf.\ \citet{wolfert2023listening}).
Furthermore, both matched and mismatched motion stimuli here have their starting points aligned to the start of a phrase in the speech, meaning that the motion in the stimulus videos might initially be more similar to each other than if the mismatched motion had been excerpted completely at random and not aligned to the start of phrase boundaries.
%Seen in this light, the 75\% preference for matched motion (after splitting ties) in FNA/UNA is a good result.
%The effect size of +25\% is half of the theoretical maximum.
It is therefore not surprising to find that the preference for matched motion over mismatched motion is not larger for FNA/UNA.

In line with expectations, no system has a relative preference for matched motion below 50\%, which is the theoretical bottom line, attained by a system whose motion has no relation to the speech.
However, the synthetic conditions are all far behind natural human motion in terms of appropriateness.
The measured effect sizes over the 50/50 bottom line range from +10\% and down to 1.5\% for all these conditions, compared to +25\% for FNA/UNA, and all differences compared to FNA/UNA are highly statistically significant.
This is a very substantial gap, and it is clear that generating meaningful and appropriate gestures is still far from a solved problem.

One other interesting trend is that a few conditions with relatively poor human-likeness, specifically FBT, UBT, and USL, show a noticeably larger fraction of tied responses, compared to other conditions.
We hypothesise that this could be due to underarticulated motion, noting that a hypothetical, extremely underarticulated system that does not move at all should receive the response ``They are equal'' all the time.
This hypothesis is consistent with the fact that these conditions all had the three lowest average acceleration values in Table \ref{tab:obj_stats}, indicating little motion overall.

\subsubsection{Comparison to the human-likeness studies}
\label{sssec:discussappvshumlike}
Compared to the results for the human-likeness studies, we did not find as many differences between the submissions in terms of appropriateness.
%, and no system came close to the performance of natural motion capture.
We can envision four factors that could contribute to this, which we list below, along with thoughts regarding potential mitigations:
\begin{itemize}
\item Responses are confined to much fewer categories, meaning that each response provides less information in an information-theoretic sense. This could potentially be addressed by having test-takers complement their response with an indication of the strength of their preference.
%That would be similar to what is done in standards of audio testing that are designed to be especially sensitive to small differences \cite{itu2015methods}, relative to the HEMVIP methodology and the standards that underpin it \cite{jonell2021hemvip,itu2015method}.
We recommend that future developments in evaluation consider using a preference scale with more response options, e.g., five or seven possible responses.
\rev{After the subjective evaluations described in this article concluded, such a scale was subsequently implemented by \citet{mehta2023diff} and \citet{kucherenko2023genea}.}
\item Unlike the HEMVIP-based human-likeness studies, the responses to the appropriateness studies were not analysed using pairwise statistical tests to control for variation between subjects and stimuli. This might have led to reduced resolving power. It might be possible to improve on the statistical analysis using, e.g., \rev{statistical} models \rev{that} account for the effects of different test takers and different videos, or by changing the study setup to allow for pairwise statistical testing. One can furthermore gather more responses per condition, which we recommend in case the same \rev{statistical analysis} methodology is used.
\item Assessing appropriateness may be a more difficult task for humans than assessing human-likeness \rev{(where test takers assessed only motion in isolation, without any associated speech audio),} meaning that there is more random variation in the responses relative to the human-likeness studies. In a signal-to-noise analogy, this means that the noise is higher. Mitigating this would probably require changing the evaluation and its task. For example, differences might become more obvious if segments were mismatched completely randomly, such that speech sometimes would be paired with motion from a segment where the character is not actively speaking, and vice versa \rev{(see \citet{wolfert2023listening})}, although doing so would essentially change the type of appropriateness that is being assessed.
\item It may simply be that current artificial systems struggle to generate motion that is particularly appropriate to any specific input speech. In other words, in a signal-to-noise analogy, the signal is weaker. Consequently, there is less of a difference to be uncovered in the first place.
\end{itemize}
Although all of these factors may contribute to the results we observe, the big gap in effect size between natural motion capture and synthetic motion,
%(+25\% for natural motion versus +10\% or less for synthetic motion),
and the fact that FNA/UNA were very significantly better than all other conditions, shows that our methodology is sufficiently accurate to clearly resolve important differences between conditions.
\rev{Coupled with the finding that FSA/USQ were significantly differently better than FNA/UNA when instead rating human-likeness, it is clear that our evaluations have managed to tap into and estimate different aspects of motion.}
%From Fig.\ \ref{fig:joint}, we can furthermore see that
%, unlike the GENEA Challenge 2020,
%there is no strong correlation between the human-likeness ratings and the appropriateness ratings in the evaluation.
%This supports a conclusion that our evaluation successfully disentangled these two aspects of gesture motion.
%This supports an interpretation that -- in contrast to human-likeness -- data-driven co-speech gesture generation still has a long way to go to come close to the speech appropriateness and specificity that human gesticulation has.
\begin{revision}

\begin{figure*}[t!]
\centering%
%\hfill
  \begin{subfigure}[b]{0.49\textwidth}
    \centering%
    \includegraphics[width=\textwidth]{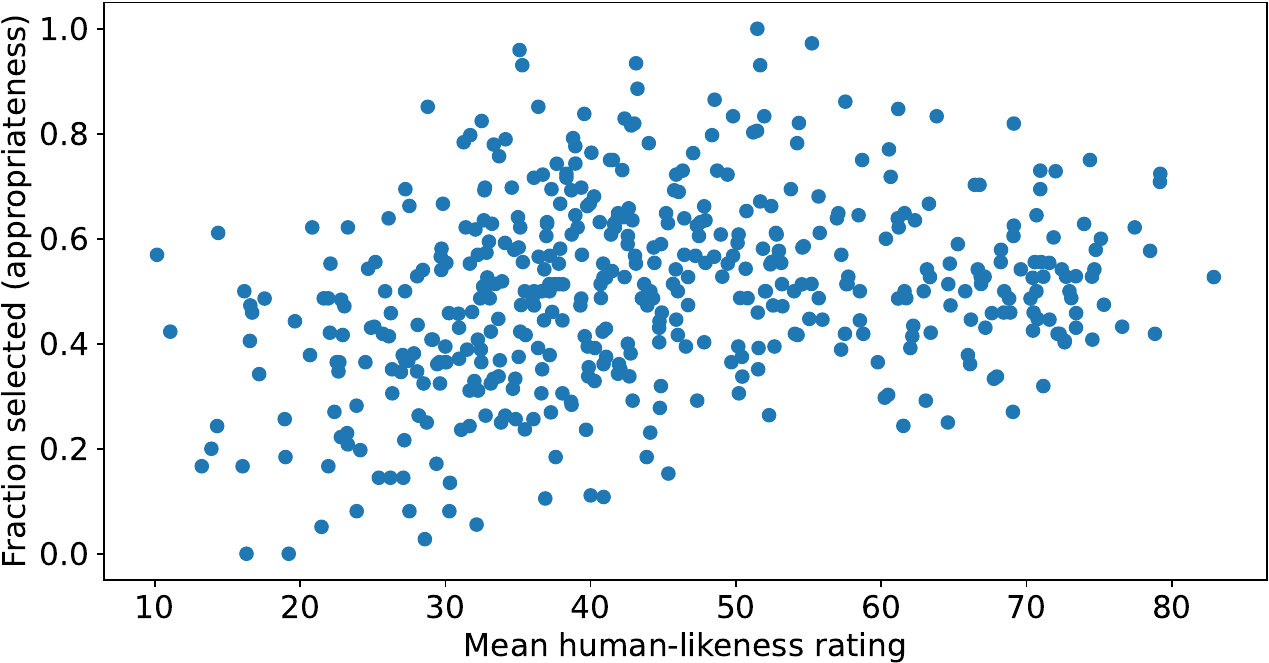}
    \caption{Full-body}
    \label{sfig:fbhlappscatter}
  \end{subfigure}
\hfill
  \begin{subfigure}[b]{0.49\textwidth}
    \centering%
    \includegraphics[width=\textwidth]{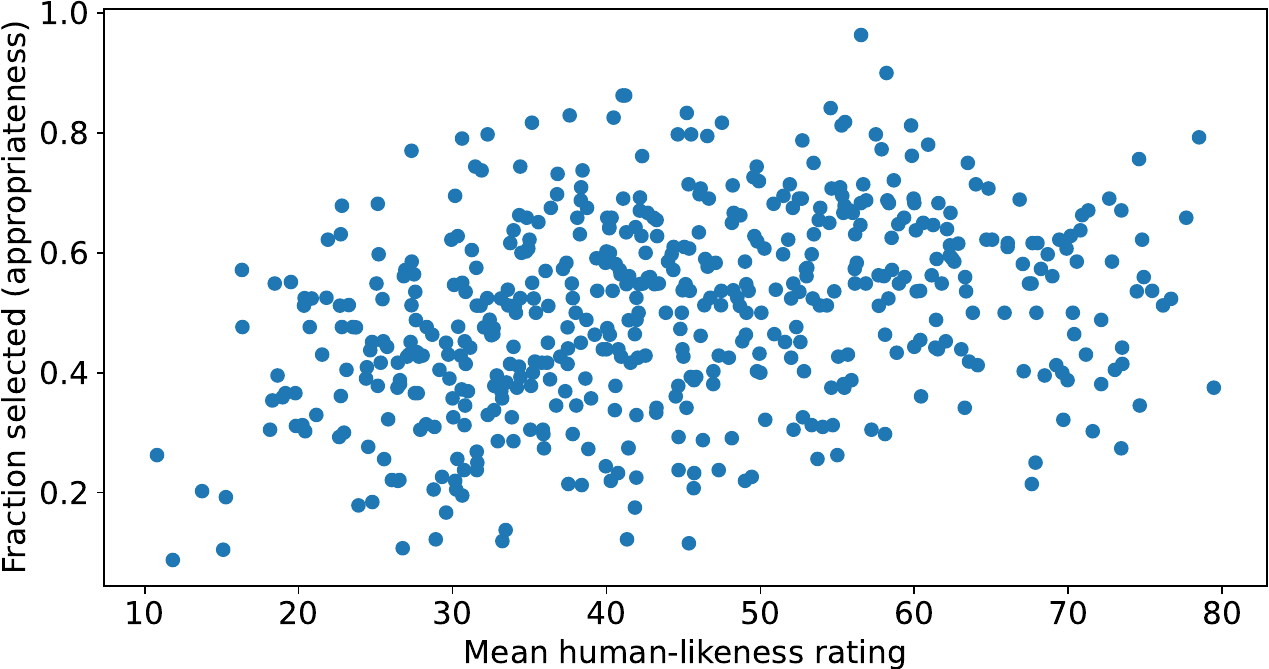}
    \caption{Upper-body}
    \label{sfig:ubhlappscatter}
  \end{subfigure}
%\hfill
\caption{\rev{Scatterplots illustrating the correlation between the rated mean human-likeness of a motion segment and how often that segment was selected by users in the appropriateness user study as being the more appropriate one -- regardless of whether it was shown as the matched or mismatched motion. Each point represents a unique motion segment. Pearson correlation analyses reveal positive correlations of 0.26 ($p$-value < 0.01) and 0.32 ($p$-value < 0.01) for full-body and upper-body motion tiers, respectively.}}
\label{fig:scatterhlapp}
\end{figure*}

\subsubsection{Analysis of stimulus-level correlation between performance measures}
\label{sssec:correlation}
To quantify the degree of decoupling between the human-likeness ratings and appropriateness assessments, we ran a numerical analysis of the correlation at the stimulus (motion segment/excerpt) level. 
%Additionally, we fitted a mixed effects model.
%(Models cannot be fitted to responses directly since the human-likeness and appropriateness evaluations did not have the same participants.)
For each motion segment from the human-likeness study, we computed its (arithmetic) mean human-likeness rating, along with how often that motion segment was chosen as the more appropriate motion example when presented in the appropriateness study -- regardless of whether it was the segment that matched the speech audio or not.
Tied responses were split equally between both stimuli in the pair.
%Instead we ran a Pearson correlation analysis on averaged human-likeness ratings and averaged appropriateness responses, with each matching response coded as $1$ and each mismatched response coded as $0$.
%Ties were coded as $0.5$.

Scatterplots visualising the resulting data for the two tiers of the challenge can be found in Fig.\ \ref{fig:scatterhlapp}.
Computing the Pearson correlation between the two quantities in the scatterplots (rated human likeness vs.\ empirical user preference in the appropriateness study) yielded a correlation of 0.26 ($p$-value <0.01) and 0.32 ($p$-value <0.01) for full-body and upper-body motion, respectively.
%For reference, the Pearson correlation between the human-likeness and appropriateness ratings (which are both continuous) in the 2020 challenge data was XXX ($p$-value XXX).
The correlation analyses thus find that a higher mean human-likeness rating is on average associated with an greater probability of a segment being selected by users in the appropriateness study, regardless of whether or not it matched the speech.

However, although statistically significant, correlations were of moderate strength.
To further control for the effect of human-likeness in the mismatching paradigm, we propose that future studies may (1) explicitly ask test takers to ignore visual motion quality in making their judgements (similar to the question formulation used in 2020 \citep{kucherenko2021large}), and/or (2) may choose to set up studies such that the matched and mismatched segments in each presented pairing have similar mean human-likeness ratings.
%Looking at the joint plot of the two performance measures in Fig.\ \ref{fig:joint} suggests that a large fraction of this correlation likely is due to FNA/UNA, which scored well on both metrics, whereas most remaining conditions achieved more moderate scores on both.
%
%The observed correlation is unlikely to be attributable a rater preference for better-looking motion, since the same motion clips were both for matched an mismatched stimuli, just permuted.
%Instead, our finding suggests that the underlying concepts are linked.
%This notion is supported by considering a hypothetical model that does not move at all, which essentially has bottom-line human-likeness, and also does not enable any discrimination between appropriate or inappropriate motion.
%Our experiments support this reasoning, in that the models with the lowest acceleration (FBT, UBT, and USL) were assigned low human-likeness and also gave rise to a large number of tied responses.
%(see Sec.\ \ref{ssec:appropriatenesscomments}).
\end{revision}

\subsubsection{Comparison to other gesture-appropriateness assessments}
The distribution of the three different  responses across the different conditions in Fig.\ \ref{fig:appropbars} is similar to that seen in the mismatching study reported in \citet{jonell2020let}, which used a similar methodology.
On the other hand, we see fewer statistical differences compared to the appropriateness study in GENEA 2020 \citep{kucherenko2021large}, which asked participants to rate the appropriateness of the stimuli on an absolute scale using HEMVIP.
However, the ratings in that study were strongly biased towards conditions with high human-likeness, as discussed in Sec.\ \ref{sec:approp}.
\rev{This is} evidenced by the fact that mismatched natural motion (M) scored second best in terms of appropriateness there.
\rev{In the new appropriateness evaluation paradigm, M would perform at chance rate by definition.
Furthermore, in a segment-level re-analysis analogous to those in Sec.\ \ref{sssec:correlation}, the Pearson correlation between the 2020 mean human-likeness and mean appropriateness ratings is 0.51, which is both significantly different from zero ($p$-value < 0.01) and numerically about twice as large as the correlations between human-likeness and appropriateness judgements in the 2022 evaluations.
The reduced correlation in 2022 indicates that responses in the latest appropriateness studies are markedly less confounded by segment human-likeness, as was our goal.}
In effect, we traded the high-resolution, high-bias method from GENEA 2020 for a reduced-resolution, \rev{lower}-bias method.
%We think this is a step forward, since prior evaluations of gesture appropriateness for speech have been \rev{substantially} confounded by motion quality, whereas our methodology is \rev{explicitly designed to control for it}.
%, in such a way that systems whose gestures that are independent of the input speech (e.g., condition M above) attain chance definition.
%The fact that some synthetic conditions that distinguished themselves the most in terms of appropriateness, namely FSH and USM, exhibited middle-of-the-pack human-likeness, highlights our success in disentangling motion appropriateness from motion quality.

%The methodology we demonstrate also has other advantages, since it does not involve subjects making any direct comparisons between videos generated by different conditions, compared to the method used by \citet{kucherenko2021large}.
In addition to \rev{controlling} for the effect of motion quality, our method for assessing appropriateness only requires comparing a system to itself.
We believe this feature may enable direct comparison between different studies on the same data, \emph{without} having to include the various other synthetic baseline conditions in the new user study.
Seeing that creating appropriate baseline systems is one of the sticking points both for carrying out research and for its subsequent assessment in peer review, this can be a major simplification compared to parallel methodologies like HEMVIP \cite{jonell2021hemvip} that involve simultaneously comparing and evaluating many different conditions against each other.
Since responses in those studies are affected by what other videos are shown on the same page, their results thus cannot be directly compared unless stimuli or implementations of previous synthetic baseline conditions are included in the new study.
%
%\subsection{Miscellaneous remarks}
%In addition to its strong ability to control for the effect of motion quality, our new method for assessing appropriateness only requires comparing a system to itself.
%This makes it easy to use and track progress on different sets of stimuli without having to train any baseline systems for the comparisons. This could be advantageous for future benchmarking purposes, since creating appropriate baseline systems is one of the sticking points both for carrying out research and for its subsequent assessment in peer review.
Our recommendation for future research that uses the same methodology \rev{as} this paper is to report effect size and $\alpha=0.05$ Clopper-Pearson confidence intervals similar to Table\ \ref{tab:stats}, to enable easy and accurate comparison between studies.

\subsection{Discussion of objective metrics}
\label{ssec:obj_discussion}
The values of each of the six objective metrics in Table \ref{tab:obj_stats} span a wide range.
From the acceleration and jerk values, we can observe that some systems, e.g., the text-based baselines from \citet{yoon2019robots}, exhibit much less movement than others.
Unfortunately, most objective metrics are not well aligned with subjective human-likeness scores. 
In the full-body tier, one of the least human-like systems, FSB, received some of the best scores in terms of average absolute jerk, acceleration, and Hellinger distance.
At the same time, one of the most human-like systems, FSC, is not in the top three according to any of the objective metrics used.
In the upper-body tier, one of the least human-like systems, USP, was in the top three systems according to average jerk, acceleration, and Hellinger distance while one of the most human-like systems, USO, is not in the top three according to any of the objective metrics.
The rank correlations in Table \ref{tab:rankcorr} make these observations more precise, by showing that most correlations are not statistically significantly different from zero.
The one exception is the FGD.
Although the correlations we found there are moderate (around $-0.5$) and system USN shows an outlying value, this metric might have some potential as an objective evaluation metric useful for faster evaluation in the development phase, although it is not clear how well it will resolve smaller differences between systems.
\rev{Further development and validation of objective metrics would benefit the research community, as exemplified by a recent study that improved the FGD and generalised it to non-human motion \citep{maiorca2023objective}.}

As for speech appropriateness, only the CCA metric takes reference motion into account and thus has any possibility to measure this aspect. \rev{(None of the studied metrics explicitly considers information from the speech itself.)}
The CCA results are not clear-cut, but nonetheless somewhat encouraging, seeing that the systems with the best appropriateness (namely FSH, USQ, and USM) also exhibit some of the highest CCA values, of 0.96 and above, and we found a statistically significant correlation for one of the tiers.

All in all, we want to emphasise that objective evaluation of generated gestures
%assessing gesture generation through objective evaluation is still challenging.
is still an open problem.
Subjective evaluation, as used by this challenge, remains the gold standard for comparing gesture-generation models \cite{wolfert2021review}, and none of the objective evaluation metrics can replace subjective user studies.

\subsection{Take-home messages}
\label{ssec:whatlearnt}
In this section we combine salient points from our findings with information that the teams provided about their challenge entries, in order to see what we can learn about what aspects matters most in gesture-generation methods, data processing, and evaluation.
%In this section we discuss what we have learnt about gesture generation modeling and evaluation as well as how to use the materials derived during this challenge.

\begin{revision}
\subsubsection{What have we learnt about the gesture-generation problem?}
The challenge results -- with some entries performing very well in human-likeness, but none coming close to human-level appropriateness -- indicate that generating random generic gesturing movements is much easier than tailoring gestures to fit the speech well.
This could be due to the fact that there is a strong correlation between consecutive frames of movement, whilst the correlation between speech and gestures is relatively weak.
One simple argument that the latter correlation is weak is that the same speech may be accompanied by different gestures, and the same gesture conversely may accompany different speech audio clips.
Furthermore, fastText \cite{bojanowski2017enriching}, the most commonly used text representation among challenge entries, embeds each word individually regardless of context, and is far from the state of the art in language modelling and text representation; cf.\ \citet{wang2019superglue}.
(Indeed, it is arguably weaker than some of the text representations used in 2020 and referenced in Sec. \ref{ssec:2020challenge}.)
This likely impeded the ability of the models to learn semantically appropriate gestures.
\end{revision}

\subsubsection{What have we learnt about successful gesture-generation methods?}
\label{subsec:system_analysis}

\rev{Sec.\ \ref{sec:systems} provides an overview of the submitted systems, and} Table \ref{tab:conditions} lists all submissions with their corresponding system properties, sorting them according to their human-likeness scores. 
We can note that all systems except the text-based baseline used audio as an input modality, fewer systems used text, and even fewer used speaker IDs.
There seems to be no clear indication that using any given combination of modalities necessarily gives better results than others, as some systems using only audio are on the top and others on the bottom of the list.
However, the fact that so many of them did use audio input suggests a perception among teams that taking audio into account is important.
\rev{This is reinforced by the finding that BT, the only system exclusively based on text, did not perform well in the subjective evaluations.}

When it comes to the techniques used, RNNs were the most popular choice and used almost by all the systems, followed closely by auto-regression.
Again, for most of these there seems to be no strong indication that certain choices are necessarily better than others.
\rev{This is not unexpected, given the many aspects and ways in which challenge submissions differ, all of which are likely to have affected the results in different ways.
Even by aggregating results across multiple challenges, associating outcomes with individual design decisions remains difficult; cf.\ \citet{king2014measuring}.}

Our main \rev{and clear} observation is that the state-of-the-art in human-likeness is not to use deep learning for everything (or at least not to generate the gesture poses), seeing that the most human-like system, GestureMaster, is based on motion graphs \citep{lee2002interactive,kovar2002motion,arikan2002interactive} and a library of carefully selected high-quality motion segments.
\rev{For methods that used deep learning to generate output poses (i.e., all other submissions), the two approaches demonstrating the greatest human-likeness in both tiers relied on probabilistic approaches (VAEs or VQ-VAEs) with stochastic output generation.
This resembles the state of the previous challenge, where the entry with the greatest human-likeness \citep{simon_alexanderson_2020_4088600} was based on pose sequences sampled from an autoregressive normalising flow, although the difference to the most human-like non-probabilistic submission \citep{vladislav_korzun_2020_4088609} was not statistically significant at the more stringent $\alpha=0.01$ level \citep{kucherenko2021large}.}

\rev{Although perhaps initially surprising, the strong showing from a playback-based method echoes the long dominance of concatenative (i.e., exemplar-based) systems in terms of speech-synthesis naturalness \citep{king2014measuring}.
Furthermore, the use of motion graphs also resembles the prevailing approach to achieving high visual quality in 3D rendering, which is to combine small constituent images (bitmaps) as textures on a mesh; pure machine-learning approaches to rendering have taken a very long time to become competitive in terms of graphical quality \citep{mildenhall2021nerf}.
In general, it appears that -- unless one has the methods and data needed to build an exceptionally strong deep generative model -- an approach based on concatenating shards of real-world observations is instead the better path to achieving convincing results.
In such an approach, it is feasible to ensure that all individual units hold high quality (they may for example be taken wholesale from the real world, making them completely natural by definition), leaving only the task of joining them together with minimal artefacts.}

\rev{However, it is not only the best submissions that have gotten better in the recent challenge.
The baselines from the 2020 challenge both performed relatively poorly in terms of human-likeness in 2022.
%The same holds true for the only team that participated in both challenges (FineMotion), which ranked much better in 2020 than in 2022 (although they used different systems for the different years).
Together, these findings offer evidence that the quality of gestures in the challenge as a whole is increasing, which may indicate that gesture-generation methodologies in general are getting better.}
\begin{figure}[!t]
    \centering%
    \includegraphics[width=\columnwidth]{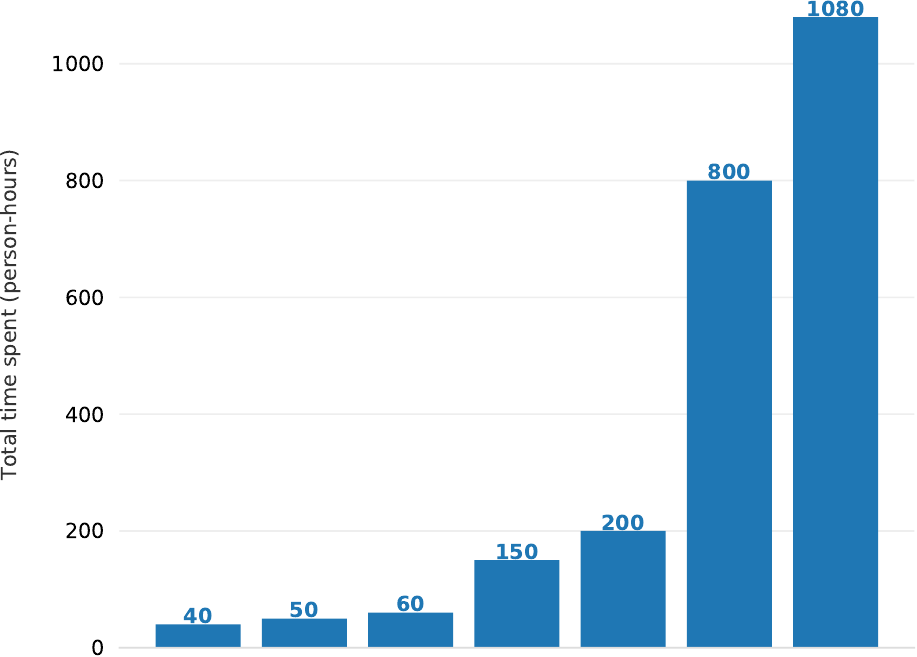}
    \caption{The number of person hours different responding teams reported spending on the GENEA Challenge 2022, sorted in ascending order.}
    \label{fig:time_spent}
    \Description{The figure shows a bar chart of the total person hours each team reported spending on the challenge. The heights of the bars correspond to 40, 50, 60, 150, 200, 800, and 1080 hours. It is not shown which team gave which response.}
    \vspace{-1\baselineskip}
\end{figure}

\subsubsection{What have we learnt about gesture-data processing?}

Fig.\ \ref{fig:time_spent} shows how much time different teams spent on their submissions.
We can see a very high variation, with some teams spending between 40 and 60 person hours whilst some others spent 800 hours or more.
The two teams who spent 800 or more hours on their submissions reported devoting a large amount of time on data pre-processing, which other teams did not.
%(although several other teams did state that they found data processing to be of significant importance).
One of the former teams is the top-performing team in terms of gesture human-likeness scores.
This suggests that spending time on data preparation is likely to pay off in better model performance\rev{, which is consistent with trends from the Blizzard Challenges in text-to-speech, where top teams often spend significant resources on manual data acquisition and processing}.
Data processing tasks included cropping the recordings into shorter segments, annotating those short segments for, e.g., motion quality, and similar.
%removing segments with silent pieces in the audio, etc.
Some teams found it important to remove segments where the character was listening rather than talking, since the character exhibits little gesture motion in these segments, which can make deterministic gesture-generation approaches regress towards the mean pose and thus produce less vivid movement.

\begin{revision}
There were variations in how challenge teams represented audio and motion in their entries, but we did not find strong evidence that certain representations were better than others.
Systems that used modern, learnt audio representations such as WavLM \citep{chen2022wavlm} and PASE+ \citep{ravanelli2020multi} (not seen in the 2020 challenge) did not show superior performance compared to systems that used conventional MFCC audio features.
For motion, submissions likewise exhibited a more diverse set of representation approaches than what was seen in 2020.
In this challenge, there was a weak finding that systems using motion representations based on rotation matrices, including 6D representations \citep{zhou2019continuity} or 2-axis rotation matrices \citep{zhang2018mode}, obtained better human-likeness scores than systems that used exponential maps \citep{grassia1998practical} of axis-angle representations.
However, this finding is not conclusive due to the small number of examples and the multitude of factors affecting system performance, and might simply reflect the fact that systems with less time put into their development were more likely to use the data pipeline and motion representation of the provided baseline code, which used exponential map representations.
\end{revision}

Another important aspect when it comes to the data is post-processing, such as hip-centering or smoothing (cf.\ \citet{kucherenko2021moving})\rev{, of the output motion}.
As seen in Table \ref{tab:conditions}, most of the systems (good and bad performance alike) applied motion smoothing in some form.
This suggests that they found smoothing to be beneficial for gesture generation, although the user studies do not allow us to make a statistical conclusion about the importance of smoothing the output motion.

Finally, modelling the motion of the fingers or having them fixed emerged as another important decision.
Roughly half of the systems in the evaluation used fixed fingers.
Some of these systems achieved good performance whilst others did not.
This does not allow us to make strong statistical conclusions about the importance of modelling fingers.
However, we may surmise that finger motion may be especially difficult to make natural, otherwise all teams would presumably have included finger motion in their submissions.
%System FSA/USQ \cite{zhou2022gesturemaster} used a quite involved procedure for selecting which finger motion to use separately from the rest of the body.
%This situation can be compared to, say, lower-body motion in the full-body tier, where all teams chose to use a moving rather than fixed lower body.

\subsubsection{What have we learnt about evaluating gesture generation?}

Previous work shows it is not easy to disentangle perceived human-likeness from appropriateness as more human-like systems are often ranked as more appropriate \cite{kucherenko2021large}.
In this challenge we made a concerted effort to disentangle these two aspects.
Specifically, we (1) muted the audio during the human-likeness evaluation, to remove any influence speech may exert on perceived appropriateness (cf.\ \citet{jonell2020let}), and (2) compared each model with a mismatched version of itself (having similar human-likeness),
%\rev{, both displaying the exact same speech audio},
to control for the effect of human-likeness when evaluating appropriateness.
This effort paid off, \rev{seeing that different conditions performed best on the two performance measures, with differences being statistically significant, whilst simultaneously ensuring that a speech-independent system (like condition M in 2020) no longer can score better than chance.}
%since the two metrics are essentially uncorrelated for the synthetic conditions in Fig.\ \ref{fig:joint}.
However, improving the statistical resolution of the evaluation procedure would be beneficial.
%Objective metrics should not replace subjective evaluation.
%This suggests a way to disentangle appropriateness and human-likeness.

% Moved the paragraph below to the new "Limitations" subsection

% We considered only the general appropriateness of the gestures for the speech, while there is value in separately evaluating appropriateness towards the speech rhythm and meaning since those are different aspects. We will consider doing that in future challenges, for example by doing two user studies each focused on a separate type of appropriateness: semantic appropriateness and rhythmic appropriateness. %, such as interlocutor speech, stance, behaviour, or gaze.

\subsection{How materials from the challenge can be used}
\label{ssec:materials}
We believe the materials released together with the challenge have many benefits for gesture-generation  research.
To illustrate this, we provide a list of possible use cases, often with references to prior work similarly that re-used resources from the previous GENEA Challenge (from 2020) in a similar manner.
%acting as examples of this kind of data use.
One may, for instance\ldots{}
%The various materials from the challenges may be used to\ldots{}
%All the material derived during the Challenge can be found at  \href{https://youngwoo-yoon.github.io/GENEAchallenge2022/}{youngwoo-yoon.github.io/GENEAchallenge2022}. 
%Here is the list of the different ways those materials can be used in the future research:%, similarly how it was done with the data from the GENEA Challenge 2020:
\begin{itemize}
    \item Benchmark/compare new models to the state of the art using our public data and existing motion or video stimuli, like \citet{ferstl2021expressgesture, yazdian2022gesture2vec} did with previous open stimuli.
    \item Evaluate models using our open-sourced code for the evaluation interface and analyses, similar to the re-use of HEMVIP code from \citet{jonell2021hemvip} by \citet{wolfert2021rate}.
    \item Use our questions and evaluation structure for evaluating new proposed methods, similar to how \citet{teshima2022deep} re-used previous evaluation designs.% same questions in the subjective evaluations,
    %as in \cite{teshima2022deep},
    %with similar approaches to disentangling human-likeness and speech appropriateness \cite{jonell2020let,rebol2021passing}.
    \item Use our public visualisation avatar and/or code to simplify development and obtain more standardised and comparable visuals, as done by \citet{Alexanderson_2023}, and similar to prior re-use of the GENEA Challenge 2020 open upper-body visualisation in \citet{saund2021importance, wang2021integrated, teshima2022deep, zhang2023diffmotion,mehta2023diff}.%, Alexanderson_2023}.
    %from the GENEA Visualiser%, as was done previously for the GENEA Challenge 2020 \cite{wang2021integrated, teshima2022deep}
    %\item Use the GENEA Visualizer for faster development and continual comparisons with the same visualization
    \item Evaluate new models objectively using the same metrics that showed the most promise here, similar to how \citet{ahuja2022low,liang2022seeg,ye2022audio} re-used metrics and sometimes code from \citet{yoon2020speech,ahuja2020no}.% as \cite{bhattacharya2021speech2affectivegestures} did
    %\item Learn a mapping from 3d sequences to subjective evaluation as in \cite{he2022automatic}
    \item Use our large dataset of subjective evaluation responses to build and/or validate new automatic quality-assessment methods, similar to \citet{he2022automatic}, or perform in-depth analyses of human preferences using the individual response data, perhaps linking these to the time taken by study participants, their questionnaire responses, etc.
    \item Use our materials and those released by participating teams to probe reproducibility in the field.
\end{itemize}

\section{Limitations}
\label{sec:limitations}
Despite being a large evaluation with many conditions and raters, there are inevitable limitations to the challenge and its results, imposed by scope, systems, data, visualisation, and evaluation choices.
We discuss some of these limitations below.

\setcounter{subsection}{1}
\subsubsection{Scope and scale}
The ten teams participating in the 2022 challenge do not represent the full spectrum of all gesture-generation approaches available today.
Although ten teams (plus the top line and baselines) are more systems than considered in any other joint comparison of gesture-generation systems we are aware of, it is still not large enough to, e.g., make strong conclusions regarding which system architectures to prefer.
We hope to attract more teams to participate in the challenge in future years.

\subsubsection{Data}
Motion capture is a remarkable technology, but does not yet perfectly capture every aspect of human pose and figure.
There are hardware issues such as calibration, and software challenges in estimating poses of diverse humanoid skeletons whilst dealing with problems like reflective marker displacements, occlusion, and markers in close proximity.
%(especially for the fingers).
%Motion capture software is at a remarkable stage in its ability to capture very good-looking animation.
%However, the technology is a complex system which is far from perfect, as a result of which errors are bound to manifest during both the data recording and processing stages.
%On the hardware side, one should consider imperfections in the equipment (i.e., cameras, reflective markers, body suits), as well as hardware calibration and changes in environment (e.g., heat, physical displacement).
%On the software side, algorithms have to make various assumptions in order to estimate a more general humanoid skeleton.
%For example, it has to somehow handle skeletons of various dimensions, imperfections in the equipment used (e.g., reflective marker displacements), and resolving situations when a marker may be occluded (particularly for the fingers).
Together, these issues may lead to problems with the data, commonly seen as artefacts in the produced motions (e.g., twitching or unnatural bone rotations), especially in the fingers.
%which may be especially pronounced in the fingers.
Although we have worked to exclude low-quality parts of the data and process it to make it more amenable to deep learning, some artefacts still remain.
We suspect that this is an important reason why generated motion could surpass the notionally natural motion capture in terms of human-likeness.
More, and more high-quality, motion data might allow for generating better gesture motion and constitute a stronger top line.
%Although we have cleaned the dataset of such occurrences, there are still some motion artefacts present (as well as missing speech audio in some parts).

Some useful information is also missing from the current data.
On the verbal side, this includes speech information removed for anonymisation.
On the non-verbal side, one prominent missing aspect is facial data, which is an important communication channel but was not recorded in the current dataset.
Neither was body form, such as muscle mass, body fat, skin, nor how these deform when muscles flex and extend, since the data has abstracted the humanoid form down to only a skeletal hierarchy.
%We furthermore separated the captured dyads into isolated speakers, to reduce the complexity of the present challenge.
Future challenges should maintain awareness of new datasets being published, and their data quality and modalities captured.
One modality worth investigating further is face motion, as it may help systems learn more appropriate gestures that relate to facial expressions and emotions.

\subsubsection{Visualisation}
The gesture visualisation used in the challenge has several limitations.
Some are dictated by the data, and some are deliberate choices to, e.g., reduce complexity.
The result is a virtual character that, whilst representative of typical gesture-generation visualisations, lacks both skin deformations and many human communication channels, such as gaze, facial expression, and lip motion.
Whilst the absence of such features can help focus attention on the body motion currently being studied, it does also lead to a less human-like character appearance overall.
Our evaluation also deliberately obscured some aspects of motion, e.g., by cropping the view so as to not show potential foot sliding and (for the upper-body tier) fixing the legs of the virtual character, which is innately unnatural.
%The visualisation of gesture motion is another area in which errors can occur.
%As a start, while form is captured in the virtual character due to it being a 3D mesh, it still does not capture skin deformations, such as due to muscle flexing and extension.
%Furthermore, several communication channels are missing.
%On the intra-level, the character is missing gaze, facial expressions, and lip motion, all of which may convey a significant amount of information to both the interlocutor and the user study participant.
%On the inter-level, the opposing interlocutor is missing entirely.
%This encompasses the above-mentioned communication channels, in addition to other ones such as pose, stance, and body motions, all of which may be somehow related to what the speaker is saying.
%Another point to consider is the fixed legs of the virtual character, which is innately unnatural.
%This is due to it being a post-processing step we applied in order to constrain the character to remain in the camera view at all times.
%Going further, the character is clipped from the knees below, which is also a constraining factor in terms of the communication bandwidth available to the study participant.
Future challenges should consider incorporating additional communication channels, e.g., facial features on a 3D mesh, to improve the realism of the virtual characters and their gestures.
%This would in turn include full-body animations and potentially facial features on a 3D mesh.

Aside from limitations on what agent behaviours are visualised and how, the interlocutor from the recorded conversations is missing entirely in both modelling and visualisation.
This was a deliberate choice to not increase the complexity of the challenge too much, but
%One reason for this might be that the motion originates from a dyadic conversation, but is evaluated without any contextual information about the interlocutor behaviour, e.g., their pose, stance, or motion with respect to the speaker.
the absence of such information prevents us from assessing interlocutor-dependent aspects of motion such as proxemics and behavioural alignment.
(We deliberately excluded turn taking, back channels, and listening behaviour from the subjective evaluation, since these are likely to look odd without seeing both sides of the conversation.)
Future challenges may opt to include information about both conversation parties in the evaluation, so that study participants can be interlocutor-aware in their responses.
However, any increases in complexity, whether due to adding additional inputs or output modalities, should be performed one step at a time, so that it is more clear which findings relate to which aspect of the complex problem that is gesture generation.

\subsubsection{Evaluation}
%This would be a competing explanation to the hypothesis in the previous section of artefacts in UNA due to fixing the lower-body to visualise the recorded motion.
%Other contributing factors could be that the full-body motion contains more behavioural variation, as the character now is moving their legs and changing position in the field of view, perhaps in response to the conversation partner.
%One reason for this might be that the motion originates from a dyadic conversation, but is evaluated without any contextual information about the interlocutor behaviour, e.g., their pose, stance, or motion with respect to the speaker.
%The absence of such information may complicate the assessment of the lower-body motion, which might be more closely linked to stance and proxemics in interactions compared to upper-body motion.
%On the other hand, such an effect seems like it should increase the interquartile range of ratings of FNA, relative to UNA, which is not what was observed.
Our core evaluation only sought to quantify two performance measures, namely subjective human-likeness and perceived appropriateness for the given speech.
Aspects such as gesture diversity, or generation speed and latency, were not measured.
Furthermore, we only studied the overall appropriateness of the gestures for the speech, but there is value in evaluating appropriateness with respect to the speech rhythm and speech meaning separately, since these are distinct aspects.
We hope to consider doing that in future challenges, for example by performing two user studies, each focused on a separate type of appropriateness: semantic appropriateness and rhythmic/temporal appropriateness.
A further extension would be to break this down into individual gesture categories, e.g., beat, iconic, deictic, and metaphoric gestures.

There are also many other kinds of appropriateness that can be assessed, e.g., appropriateness for the given speaker, and for the interlocutor behaviour as discussed above.
(See the discussion of \emph{grounding} in \citet{nyatsanga2023comprehensive} for a more extensive list.)
None of these were considered in the present challenge, either due to dataset limitations or to keep the complexity to a manageable level.
A difficult but important long-term goal is to pursue a more ``ecologically valid'' evaluation, to eventually compare different gesture-generation methods in human interaction, similar to \citet{he2022evaluating}.

\section{Conclusions and implications}
\label{sec:conclusion}
We have hosted the GENEA Challenge 2022 to compare many different gesture-generation methods and assess the state of the art in data-driven co-speech gesture generation for full-body and upper-body avatars.
The central design goals of the challenge were (1) to enable direct comparison between many different gesture-generation methods whilst controlling for factors of variation external to the model, namely data, embodiment, and evaluation methodology, and (2) to disentangle the effects of motion human-likeness and motion appropriateness in the evaluations.

Our evaluation results show that, with the right approach, synthetic motion can attain human-likeness ratings equal or better than the underlying motion-capture data.
This is a big step forward, although most systems did not come close to this level of performance.
The results also suggest that the field is advancing measurably, since most submissions performed significantly better than the previously published baseline methods.
However, using a careful evaluation paradigm, we find that synthetic gestures are much less appropriate for the speech than human gestures, also when controlling for differences in human-likeness.
We are thus only at the beginning of the road when it comes to generating co-speech motion that is appropriate for the specific speech.
Finally, most objective metrics we computed did not exhibit any statistically significant correlations with our subjective human-likeness ratings, with the Fr{\'e}chet gesture distance being the lone exception to the rule.
Objective metrics should thus only be used with great caution.

%The collected results suggest that the field is advancing measurably, since most submissions performed significantly better than the baselines.
%Different systems were also found to be good at different things on the two scales (human-likeness and appropriateness) that we assessed.
%However, a substantial gap remains between synthetic and natural gesture motion, indicating that gesture generation is far from a solved problem.
%Nevertheless, these results illustrate that we currently have the means to be able to generate quite convincing data-driven 3D gesture motion (although attaining that level of quality is something that only few systems are capable of at present), but we are only at the beginning of the road when it comes to generating co-speech motion that is appropriate for the specific speech.

\subsection{Implications}
The challenge findings have implications for both research and practice.
We summarise our perspectives below.
%We attempt to summarise our perspectives on these here.

% Seems useful for a journal
\subsubsection{Implications for practical systems}
If you are building a gesture-generation system and want to reach top-of-the-line human-likeness, you should currently consider using ``playback-based'' methods
like motion graphs \citep{lee2002interactive,kovar2002motion,arikan2002interactive} 
%or like motion matching \cite{buttner2015motion}
as demonstrated by GestureMaster \cite{zhou2022gesturemaster}
to generate the pose sequences, instead of relying solely on deep learning to go all the way from input features to motion.
Playback-based systems need less data, and the quality of the motion material is then a higher priority than database size, in contrast to current deep-learning trends.
%a small but carefully vetted motion database might be more important than gathering large amounts of motion data.
Machine learning is still useful for deciding which gestures to generate (e.g., which motion clips to concatenate).
In all cases, it appears important to spend time on data processing.
%If you are building a gesture-generation system, you can reach near top-of-the-line human-likeness and an appropriateness not far from that of typical synthetic systems, simply by relying on playback of pre-recorded motion, without much (or any) regard for the speech beyond its onset and offset.
%This could save a lot of technical complexity and time, and would allow for scalability by extending the motion database with new gestures as needed (however, doing so may also entail retraining any machine-learning models used).
%It appears especially important to spend time on data processing, to make sure that the data has high quality and only contains human-like motion.

\subsubsection{Implications for research and evaluation}
We believe the challenge adds value to the research community in several ways.
A lot can doubtlessly be learnt from the system-description papers by the participating teams.
The materials we release from the challenge (e.g., time-aligned splits of audio, text, and gesture data;
visualisation; code; and evaluation stimuli and responses) have broad utility for future research, system building, and benchmarking in gesture generation, similar to the community uptake of the resources from the GENEA Challenge 2020.
%
%\subsubsection*{Evaluations}
Furthermore, the methodology we demonstrate for assessing motion appropriateness for speech is much more accurate at controlling for the effect of subjective motion quality and does not involve subjects making any direct comparisons between videos generated by different conditions, which is beneficial for efficient benchmarking against previous publications (see \rev{Secs}.\ \ref{sssec:discussappvshumlike} \rev{and \ref{sssec:correlation}} for \rev{further} details and recommendations).
%, compared to the method previous challenges used.
%We believe this may enable direct comparison between different studies on the same data, \emph{without} having to include the various other synthetic baseline conditions in the new user study.
%This is a major simplification compared to parallel methodologies like HEMVIP \cite{jonell2021hemvip}, which involve simultaneously comparing and evaluating many different conditions against each other.
%Since responses in those studies are affected by what other videos are shown on the same page, studies thus cannot be directly compared unless stimuli or implementations of previous synthetic baseline conditions are included in the new study.

\subsubsection{Implications for future developments in the field}
Based on the fact that one condition in each tier managed to achieve excellent human-likeness, we expect that, in the medium-term future, gesture-generation systems (at least ones based on motion playback)
%(at least ones leveraging methods such as motion graphs \cite{lee2002interactive,kovar2002motion,arikan2002interactive} or motion matching \cite{buttner2015motion,zhou2022gesturemaster})
should be able to advance to more consistently match, or possibly even exceed, motion capture in terms of human-likeness.
Systems that generate poses directly from deep learning are likely to improve in human-likeness as well, as larger datasets with more accurate motion become available (e.g., \citet{liu2022beat}).
This would be similar to recent developments in verbal behaviour generation, where neural language models \cite{brown2020language} and speech synthesisers \cite{shen2018natural,li2019neural} trained on large datasets are approaching the text and speech produced by humans in terms of surface quality (but not necessarily appropriateness).
Gesture generation may be lagging behind due to the relative scarcity of high-quality \rev{3D} motion data, compared to text and audio, since accurate motion estimation from monocular in-the-wild video remains a challenging problem.

As the \rev{above} evolution runs its course, we believe that research into appropriate rather than human-like motion is poised to become the new frontier in gesture generation.
There is already evidence that existing deep-learning methods in principle can predict what specific properties are appropriate for each individual gesture instance to be generated, even for the hard case of semantically motivated, communicative gestures from speech \cite{kucherenko2022multimodal, kucherenko2021speech2properties2gestures,ao2022rhythmic}.
We also believe that there is great potential for devising better objective metrics, using challenge materials to validate these, and that the adoption of meaningful \rev{and validated objective} metrics may further accelerate progress in the field.

\subsubsection{Implications for future challenges}
We think that future challenges should study more difficult scenarios that are farther from being solved, for example full-body motion in dyadic interaction.
That can also provide interesting opportunities for exploring other types of appropriateness, e.g., with respect to the interlocutor stance and behaviour, as studied in \citet{jonell2020let, woo2021development}.
Generating interlocutor-aware full-body gestures was therefore a focus of the GENEA Challenge 2023 \citet{kucherenko2023genea}.
This should be coupled with further method development to obtain methodologies for conducting and analysing appropriateness tests with increased resolving power whilst still controlling for motion human-likeness.
In general, challenges like the one described here can play an important part in identifying key factors for generating convincing co-speech gestures in practice, and help drive and validate future progress towards
%the goal of
endowing embodied agents with natural and appropriate gesture motion.

\begin{acks}
The authors wish to thank Meta Research for the data; Carolyn Saund, Axel Johansson, Christianne Sandstig, Leonhard Grosse, Natalia Kalyva, and Natasha Greenwood for the transcriptions; Esther Ericsson for the 3D character; Zerrin Yumak for input; Judith Bütepage, Minsu Jang, Tony Belpaeme, Jieyeon Woo, and Rajmund Nagy for feedback on the manuscript; and Jim Royal for video production.

This research was partially supported by the Industrial Fundamental Technology Development Program (no.\ 20023495) funded by MOTIE, Korea, by the Flemish Research Foundation (FWO) grant no.\ 1S95020N, by the Portuguese Foundation for Science and Technology grant no.\ SFRH/BD/127842/ 2016, and by the Knut and Alice Wallenberg Foundation, both through Wallenberg Research Arena (WARA) Media and Language -- with in-kind contribution from the Electronic Arts R\&D department, SEED -- and through the Wallenberg AI, Autonomous Systems and Software Program (WASP).
\end{acks}

\bibliographystyle{ACM-Reference-Format}
\bibliography{refs}


\end{document}